\documentclass[12pt,showpacs,eqsecnum,nofootinbib,floats,aps,prd,floatfix,titlepage,tightenlines]{revtex4} 
\usepackage{graphicx}
\usepackage{graphics}
\usepackage{bm}
\usepackage{amssymb}
\usepackage{amsmath}
\usepackage{mathrsfs}

\begin{document}
\title{Graviton creation by small scale factor oscillations in an expanding universe}

\author{Enrico D. Schiappacasse}
\email{Enrico.Schiappacasse@tufts.edu}
\author{L. H. Ford}
\email{ford@cosmos.phy.tufts.edu}
\affiliation{Institute of Cosmology, Department of Physics and Astronomy\\
Tufts University, Medford, Massachusetts 02155, USA
\vskip 0.5in}

\begin{abstract}
We treat quantum creation of gravitons by small scale factor oscillations around the average of an expanding universe. 
Such oscillations can arise in standard general relativity due to oscillations of a homogeneous, minimally coupled scalar field.
They can also arise in modified gravity theories with a term proportional to the square of the Ricci scalar in the
gravitational action.  The graviton wave equation is different in the two cases, leading to somewhat different creation rates.
Both cases are treated using a perturbative method due to Birrell and Davies, involving an expansion in a conformal coupling
parameter to calculate the number density and energy density of the created gravitons. Cosmological constraints on
the present graviton energy density and the dimensionless amplitude of the oscillations are discussed.
 We also discuss decoherence of quantum systems produced by the spacetime geometry
fluctuations due to such a graviton bath. 
\end{abstract}


\maketitle

\section{Introduction}

A time-dependent spacetime metric can result in quantum particle creation, as was first discussed by Parker~\cite{Parker}
in the context of the expansion of the universe. The cosmological creation of gravitons was discussed by 
Grishchuk~\cite{Grishchuk:1975}, using the equation for tensor perturbations of an expanding universe found by Lifshitz~\cite{Lifshitz}. 
The process of quantum particle creation has been studied subsequently in the context of inflation. After the end of inflation,  quantum 
creation of  particles, including gravitons, can contribute to the matter and radiation of the universe~\cite{Ford:1987}. 
We here focus on a different scenario involving graviton production due to rapid oscillations
around a mean expansion rate in a spatially flat Friedmann-Robertson-Walker (FRW) background.  We consider two cosmological 
models in which these kinds of oscillations arise. The first one involves the usual matter fields in standard general relativity plus a 
minimally coupled scalar field (GRSF) in a harmonic potential.  The second model involves $f(R)$ gravity, when a term 
proportional to the square of the Ricci scalar is added to the Einstein-Hilbert action, and can arise in semiclassical 
gravity coupled to the renormalized  expectation value of a quantum matter stress tensor. Although both 
models lead to  quantum graviton creation, the graviton wave equation, which determines the creation rates, is different for 
each case. The framework of the GRSF model is standard general relativity, so the graviton equation is that obtained 
by Lifshitz~\cite{Lifshitz}, and in the transverse, tracefree gauge, has the form of the Klein-Gordon equation for a massless, 
minimally coupled scalar field.  For this reason, the problem of calculating graviton creation in the GRSF model can be reduced to 
that of calculating scalar particle production~\cite{FordParker:1977}. In the case of $f(R)$ gravity, the modified Einstein equation includes higher order derivative terms which lead to a modified  graviton wave equation~\cite{Hawng:1991, HawngNoh:1996}.  

This paper is organized as follows: In Sec.~\ref{perturbationcalculation}, we review a perturbation formalism which will be used  to 
calculate the graviton production rate. We also describe how, in both models,  an oscillating 
scale factor in a spatially flat FRW background can arise, and give explicit results for the number and energy density of the 
gravitons created by oscillations around a flat background. In Sec.~\ref{scalinggravitondensity}, we calculate 
the graviton energy density for both models in an expanding universe. In Sec.~\ref{cosmlogicalconstraints}, we discuss observational 
constraints on the energy density of the created gravitons, and hence on the oscillation amplitude of the scale factor. 
In Sec.~\ref{quantumdecoherenceinducedbygravitonbath}, we estimate the decoherence time of quantum systems induced 
by spacetime geometry fluctuations due to the  graviton bath. 
In Sec.~\ref{summarydiscussion}, we summarize and discuss our main results. 
In the Appendices, we derive in detail the oscillating scale factor and the Friedmann equation for each model.
 Units in which $\hbar = c =1$ are used 
throughout the paper.  We define the reduced Planck mass to be $M_{pl}\equiv (8\pi G)^{-1/2}$, where $G$ is Newton's
constant. The metric signature is $(-,+,+,+)$, 
Greek indices run from 0 to 3, and Latin indices for spatial components run from 1 to 3 .  

\section{Perturbation calculation of graviton creation}
\label{perturbationcalculation}

\subsection{Perturbation expansion about conformal coupling}

We take the metric to be that of a spatially flat FRW universe, with the following line element:
\begin{equation}
    ds^2=-dt^2+a^2(t)d\textbf{x}^2=a^2(\eta)(-d\eta^2+d\textbf{x}^2)   \, ,
    \label{eq:metric}
\end{equation}
where the conformal time $\eta$ is related to the scale factor $a(t)$ by $\eta=\int^{t}a^{-1}(t')dt'$. In this conformally flat spacetime, 
gravitons in general relativity, using the transverse tracefree gauge,  are equivalent  to a pair of massless minimally coupled 
scalar fields~\cite{FordParker:1977}.  Each scalar field corresponds to one of the independent polarization states of the gravitons.
In our case, we calculate scalar particle production in the metric that we are interested in, Eq.~(\ref{eq:metric}), and then multiply the final expressions for the number density and energy density of the massless scalar field by a factor of 2. (For  
discussions about graviton creation in Robertson-Walker universes, including calculations of number and energy densities, 
see Refs.~\cite{Ford:1987, FordParker:1977}.)  

The massless scalar field $\phi(x)$ satisfies the wave equation
\begin{equation}
\left[\square  - {\xi} R(x) \right]\phi(x)=0 \, ,
\label{waveequation}
\end{equation}
where $\square=\nabla_{\mu}\nabla^{\mu}$ is the covariant d'Alembert operator, $R(x)$ is the Ricci scalar, and ${\xi}$ is the coupling constant between the scalar field and scalar curvature. The minimal coupling corresponds to $\xi=0$, which is a necessary condition to study graviton production using the scalar field equation. 

In general, obtaining an exact solution of Eq.~(\ref{waveequation}) in a given metric can be difficult. We adopt an approximation 
developed by Birrell and Davies~\cite{BD80,BirrellDavies:1982}, which is a perturbation expansion about the conformally invariant case,
$\xi =1/6$.  After the mode decomposition of the field in modes $u_\mathbf{k}$, which satisfy Eq.~(\ref{waveequation}) and 
the separation of these modes as $u_\mathbf{k}(x)=(2\pi)^{-\frac{3}{2}}\exp(i{\bf{k\cdot x}})a^{-1}(\eta)\chi_k(\eta)$, the equation for 
$\chi_k(\eta)$ becomes 
\begin{equation}
\frac{d^2 \chi_k(\eta)}{d\eta^2} + \left[k^2 -V(\eta)\right]\chi_k(\eta)=0 \, . 
\label{xkequation} 
\end{equation}
Here $k =|\mathbf{k}|$ and
\begin{equation}
V(\eta)= \left(\frac{1}{6} -\xi \right)\,   a^2(\eta)\,R(\eta) \, .  
\label{vetaequation}
\end{equation}
The Ricci scalar for the spacetime of Eq.~(\ref{eq:metric}) can be expressed as
\begin{equation}
R=C^{-1}\left(3\dot D + \frac{3}{2}D^2 \right) \, , 
\label{ricciscalar}
\end{equation}
where $D=\dot{C}/C$, $C(\eta)=a^2(\eta)$, and dot denotes the derivative with respect to $\eta$. 
We impose the conditions
$V(\eta)\rightarrow0$ as $\eta\rightarrow \pm \infty$. Then the normalized  solution of Eq.~(\ref{xkequation}) which has
positive frequency in the past is denoted by $\chi_k(\eta)$, and has the asymptotic form  $\chi_k(\eta) \sim\chi_k^{in}(\eta) $,
as $\eta\rightarrow -\infty$, where 
\begin{equation}
\chi_k^{in}(\eta)=(2 k)^{-\frac{1}{2}}\exp(-i k \eta) \, . 
\end{equation}
 With this initial condition, Eq.~(\ref{xkequation}) can be replaced by an integral equation
\begin{equation}
\chi_k(\eta)=\chi_k^{in}(\eta)+k^{-1} \int_{-\infty}^{\eta} V(\eta')\,\sin\left[k(\eta-\eta')\right]\chi_k(\eta')d \eta' \, .
\label{integralequation} 
\end{equation}

The perturbation expansion results from successive iterations of this equation, and may be viewed as an expansion in
powers of $1/6 -\xi$.  We will work to  first order, and replace $\chi_k(\eta')$ by  $\chi_k^{in}(\eta')$ in the integrand 
of  Eq.~(\ref{integralequation}). The resulting solution for $\chi_k(\eta)$ may be expressed in the late time region as
\begin{equation}
\chi_k^{out}(\eta)=\alpha_k \,\chi_k^{in}(\eta)+\beta_k \,\chi_k^{in *}(\eta)\,,
\end{equation}
where the Bogoliubov coefficient, $\beta_k$, is given by
 \begin{equation}
\beta_k=-\frac{i}{2k}\int^{\infty}_{-\infty}\exp(-2i k\eta)V(\eta)d\eta \,. 
\label{betacoefficients}
\end{equation}

 The number density per unit of proper volume of created particles at late times is
\begin{equation}
n=2\times\left[2\pi^2 a^{3}(\eta)\right]^{-1}\int_0^{\infty} \vert\beta_{k}\vert^2 k^2 dk \, , 
\label{densitynumber}
\end{equation}
and the corresponding energy density is
\begin{equation}
\rho=2\times\left[2\pi^2 a^{4}(\eta)\right]^{-1}\int_0^{\infty}\vert\beta_{k}\vert^2 k^3 dk \, . 
\label{energydensity}
\end{equation}
Here the factors of $2$ account for the polarization states, and the factors of $1/a^3$ and $1/a^4$ describe the dilution and 
redshifting of massless particles by the continued
expansion of the universe after the creation process has essentially finished.

 After substituting Eqs.~(\ref{vetaequation}) and (\ref{betacoefficients}) into Eqs.~(\ref{densitynumber}) and (\ref{energydensity}), 
 and performing the respective integrals in $k$, the number and energy density can be rewritten as coordinate-space integrals, as shown
 in Refs.~\cite{BD80,BirrellDavies:1982}, 
\begin{equation}
n=2\times[16\pi a^3(\eta)]^{-1}\int^{\infty}_{-\infty}V^2(\eta_1)d\eta_1 \, ,
\label{densitynumbercoordinatespace}
\end{equation}
and
\begin{equation}
\rho=-2\times[32\pi^2a^4(\eta)]^{-1}\int^{\infty}_{-\infty}d\eta_1 \int^{\infty}_{-\infty}d\eta_2 \;
\frac{\ln\left|( \eta_2-\eta_1 ) \mu\right|^2}{2} \times \dot{V}(\eta_1)\dot{V}(\eta_2) \, .
\label{energydensitycoordinatespace}
\end{equation}
Here  $\mu$ is an arbitrary mass. The energy density $\rho$ is independent of $\mu$, provided that
$\dot{V}(\eta) \rightarrow 0$ as $\eta \rightarrow \pm \infty$. In general, the energy density of gravity waves, and hence of 
gravitons, may not be clearly defined. However, when the wavelength of the gravity waves is short compared to the
radius of curvature of the background spacetime, there is a well-defined effective energy momentum tensor for gravity,
as is discussed, for example, in Ref.~\cite{MTW}. This will be the case in the models we examine, as the period of the
scale factor oscillations is very short compared to the Hubble time of the FRW background. The graviton energy density
used here is obtained from this effective energy momentum tensor, as discussed in Ref.~\cite{FordParker:1977}.

Note that we are working to first order in a perturbation expansion in powers of $1/6$, so the lowest order results are only
approximate but should be adequate for the order of magnitude estimates which we seek.   

\subsection{Oscillating scale factors in a spatially flat FRW background}
\label{oscillatingscalefactor}

We consider small oscillations around a FRW background, with a scale factor of the form
\begin{equation}
a(t)= \bar{a}(t)\left[ 1 + A_{\text{eff}}(t) \cos(\omega_0 t) \right] \, , 
\label{scalefactorinanexpandinguniverse}
\end{equation}
where $\bar{a}(t)$ is the background scale factor time averaged over oscillations, $A_{\text{eff}}(t) \ll 1$ is a nonconstant 
oscillation amplitude, and $\omega_0$ is the angular frequency of oscillations.  Note that if we take the background scale factor 
to be that of flat spacetime and use $t \approx \eta$ to leading order, then Eq.~(\ref{scalefactorinanexpandinguniverse}) takes 
the following form in conformal time:
\begin{equation}
a(\eta)= 1 + A_0 \cos(\omega_0 \eta) \, , 
\label{scalefactor}
\end{equation}
where $A_0 \ll 1$ is the constant amplitude of the metric oscillations. 

We analyze two models in which a scale factor of the form in Eq.~(\ref{scalefactorinanexpandinguniverse}) can arise.
First, we consider the standard matter fields in general relativity consisting of a perfect fluid plus the addition of a minimally coupled 
scalar field in a harmonic potential. Second, we consider a specific model in $f(R)$ gravity in which the gravitational 
action is expanded in a power series to second order in the Ricci scalar. 

\subsubsection{Standard matter fields in general relativity plus a minimally coupled scalar field (GRSF model)}

Coherent scalar field oscillations in an expanding universe were studied by Turner \cite{Turner:1983}, and have been widely 
considered in the literature in the context of
inflation and the reheating epoch after inflation~\cite{Shtanovetal:1994} or as a dark matter 
candidate~\cite{PeeblesVilenkin:1999,SuarezMatos:2011}. We focus on the oscillations of the scale factor driven by scalar field 
oscillations.  The action for this model is given by
\begin{equation}
 S = \frac{M_{pl}^2}{2}\int d^4x \sqrt{-g}R + \int d^4x \left[ \mathscr{L}_M(g_{\mu\nu},\Psi_M) + 
 \mathscr{L}_{\text{scalar}}(g_{\mu\nu},\varphi)\right]\,,
\label{G(R)plusscalaraction}
\end{equation}
where $\mathscr{L}_M(g_{\mu\nu},\Psi_M)$ is the Lagrangian for the matter fields $\Psi_M$,  and 
$\mathscr{L}_{\text{scalar}}(g_{\mu\nu},\varphi) = (\sqrt{-g}/2)[-g^{\mu\nu}\partial_{\mu}\varphi\partial_{\nu}\varphi - 2V(\varphi)]$, 
where $\varphi$ is a homogeneous scalar field with a harmonic potential, $V(\varphi)=(\omega^2\varphi^2)/2$.  The  
Friedmann equation for the scale factor is
\begin{equation}
 3H^2M_{pl}^2  =\rho_M + \rho_{\varphi}\,,
 \label{scalar1}
 \end{equation}
 and the scalar field equation of motion is
 \begin{equation}
 \partial_{t}^2\varphi +  3H\partial_{t}\varphi + \omega^2 \varphi = 0\,.
\label{cosmologicalequationonescalar}
\end{equation}
Here  $H = \dot{a}(t)/a(t)$ is the Hubble parameter and $\rho_M$, and $\rho_{\varphi} =(\partial_t \varphi)^2/2 + V(\varphi)$ are 
the energy density for matter fields and the scalar field, respectively. In the regime $H \ll \omega$, the friction term in 
Eq.~(\ref{cosmologicalequationonescalar}) is subdominant, and the scalar field oscillates around the minimum of the potential 
with an angular frequency $\omega$ according to $\varphi(t) \approx A(t) \cos(\omega t)$. Let $A(t)\propto 1/\bar{a}(t)^{\gamma}$.
 Then, if we neglect $\ddot{\bar{a}}(t)$ and $\dot{\bar{a}}^2(t)$ terms and take $H \approx \bar{H} \equiv \dot{\bar{a}}(t)/\bar{a}(t)$, the  expression for $\varphi(t)$ 
satisfies Eq.~(\ref{cosmologicalequationonescalar}) with $\gamma = 3/2$.
It follows that the time evolution of the scalar field can be expressed as
\begin{equation}
\varphi(t) = \varphi_i\left( \frac{\bar{a}_i}{\bar{a}} \right)^{3/2} \cos(\omega t)\,,
\label{timeevolutionscalar} 
 \end{equation}
where $\varphi_i$ is the oscillation amplitude when oscillations start at time $t_i$ and $\bar{a}_i \equiv \bar{a}(t_i)$. 
The oscillating behavior of the scalar field causes  the scale factor in this model also to have an oscillating behavior. 
In  Appendix~\ref{GRscalar} we calculate this scale factor in detail, and find
\begin{equation}
a(t)=\bar{a}(t) \left[ 1 - D_i \left(\frac{\bar{a}_i}{\bar{a}(t)}\right)^{3} \cos(2\omega t) \right]\,,
\label{scalefactorGRscalar}
\end{equation}
 where $D_i\equiv (\varphi_i^2) / (16 M_{pl}^2)$ is the initial amplitude of the metric oscillations. Thus the scale factor oscillates
 at twice the frequency of the scalar field.  
If we consider this background scale factor to be that of flat spacetime and $D_i \ll 1$, Eq.~(\ref{scalefactorGRscalar}) takes 
the form, to leading order, of Eq.~(\ref{scalefactor})  with $D_i = A_0$ and $ \omega = \omega_0 / 2$, where $\omega$ is 
the mass of the scalar field $\varphi$.

The generation of gravitons in this model is ruled by Eq.~(\ref{xkequation}), because we are working in standard general relativity. 

\subsubsection{Modified Einstein's gravity: Quadratic terms in the curvature [$f(R)$ model]} 

Oscillations of the scale factor shown by Eq.~(\ref{scalefactorinanexpandinguniverse}) can also arise from modifications 
of Einstein's equation by terms quadratic in the curvature.
An example is  $f(R)$ gravity, where the Einstein-Hilbert action is taken to be $S_H=\frac{1}{2}\,  M_{pl}^2 \,\int d^4x \sqrt{-g} f(R)$,
with $f(R)$ being an analytic function of  the Ricci scalar $R$. Expand $f(R)$ to second order as
\begin{equation}
f(R)=a_0 + a_1R+\frac{a_2}{2!}R^2 +\ldots \, 
\label{ricciscalarexpansion}
\end{equation}
and set $a_0=0$ and $a_1=1$, so that $f(R) \approx R+(a_2/2)R^2$. The resulting  modified vacuum  Einstein's 
equation and its trace equation are, respectively,   
\begin{equation}
G_{\mu\nu} + a_2\, \left(RR_{\mu\nu}-\frac{1}{4}R^2g_{\mu\nu}+
g_{\mu\nu}\nabla^{\alpha}\nabla_{\alpha}R-\nabla_{\mu}\nabla_{\nu}R\right)=0 \, ,
\label{modifiedeinsteinequation} 
\end{equation}
\begin{equation}
\left( \Box - \frac{1}{3a_2} \right)R = 0\,,
\label{tracemodifiedeinsteinequation}
\end{equation}
where $R_{\mu\nu}$ is the Ricci tensor, and $G_{\mu\nu} = R_{\mu\nu}-g_{\mu\nu}\,{R}/2$ is the Einstein tensor. The term proportional to $a_2$ in Eq.~(\ref{modifiedeinsteinequation}) need not arise from a modification of the gravitational action, but perhaps more plausibly, can also arise in semiclassical gravity where the renormalized expectation value of a quantum matter stress tensor acts as the source of gravity. 
 
In either case, the modified Einstein's equation, Eq.~(\ref{modifiedeinsteinequation}), contains terms which are fourth
order in the metric and can cause flat spacetime to be unstable or to oscillate, as was discussed by
Horowitz and Wald~\cite{HorowitzWald:1978}. Let the spacetime metric be that of Eq.~(\ref{eq:metric}) with  
$a(\eta) = 1+ \gamma$. To first order in $\gamma$, Eq.~(\ref{modifiedeinsteinequation}) becomes
\begin{equation}
-\partial_{\mu}\partial_{\nu}\gamma + (\Box\gamma)\eta_{\mu\nu}+3a_2\partial_{\mu}\partial_{\nu}(\Box\gamma)-3a_2\Box(\Box\gamma)\eta_{\mu\nu}=0 \, , 
\end{equation}
where $\Box=\partial^{\alpha}\partial_{\alpha}$. The spatially homogeneous solutions of this equation grow exponentially
in $\eta$ if $a_2<0$, so flat spacetime becomes unstable. If $a_2>0$, the scale factor oscillates, as described by 
Eq.~(\ref{scalefactor}),
with an angular frequency of
\begin{equation}
\omega=\frac{1}{\sqrt{3a_2}}\;\;\;\;(a_2>0) \, .   
\label{relationomegaa2}
\end{equation}
Note the peculiar fact that as $a_2$ becomes smaller, the frequency of oscillation $\omega$ becomes larger. 
Laboratory tests of the inverse square law of gravity place an upper bound on $a_2$ of about~\cite{BerryGair:2011}
  $a_2 \alt 2\times10^{-9}\text{ m}^2$. From Eq.~(\ref{relationomegaa2}), this bound leads to a lower bound on $\omega$ of
\begin{equation}
\omega \agt \omega_B = 1.3\times 10^4\text{ m}^{-1}=4\times10^{12}\text{ Hz}. 
\label{omegalowerbound}
\end{equation}

The possible effect of these oscillations in causing radiation by charged particles was discussed by 
Horowitz and Wald~\cite{HorowitzWald:1978}, and their possible role in causing enhanced quantum fluctuation effects
through noncancellation of anticorrelated fluctuations was treated in Ref.~\cite{ParkinsonFord:2014}. Our primary
interest is their effect on graviton creation, which will be treated in the next subsection. Quantum creation 
of particles by metric oscillations plays a role in the Starobinsky model of inflation~\cite{Starobinsky:1980}, and 
was discussed by Vilenkin~\cite{Vilenkin:1985}. More recent treatments of graviton creation in oscillating metrics
have been given in Ref.~\cite{Bag:2014} in the context of emergent cosmology and in Ref.~\cite{Ema:2015} in
a model of inflaton decay.

Gravitational waves in general relativity are associated with a massless spin two graviton field with two different polarizations. 
However, the presence of higher order derivative terms in the modified Einstein's equation in  Eq.~(\ref{modifiedeinsteinequation}) 
causes, in addition to a graviton field, an extra scalar mode associated with a massive spin zero field.
To see this extra scalar mode, consider small deviations from a flat background of the form $g_{\mu\nu} = \eta_{\mu\nu} +  h_{\mu\nu}$, 
where $|h_{\mu\nu} | \ll 1$. If we work to first order in the perturbation, then the linearized version of the trace field equation, Eq.~(\ref{tracemodifiedeinsteinequation}), predicts scalar modes satisfying a massive Klein-Gordon equation~ \cite{BerryGair:2011}
\begin{equation}
\left( \Box - \omega^2 \right)R^{(1)} = 0\,,
\end{equation}  
where $R^{(1)} = \partial_{\mu}\partial_{\nu}h^{\mu\nu}-\eta^{\alpha\beta}\Box h_{\alpha\beta}$ is the linearized Ricci scalar to 
first order. Thus, if we take seriously this modified gravity theory, we should expect massive scalar particle creation together to 
graviton creation. We will focus solely upon graviton creation  in the present paper. We expect the scalar particle creation 
 rate to be somewhat suppressed compared to that for gravitons due to the nonzero mass of the scalar particle and the two
 polarization degrees of freedom of the gravitons. In any case, the observational constraints which
 we will derive using gravitons alone may be regarded as lower bounds on the slightly tighter constraints which would
 arise if the effects of scalar particles were also included.
 
 Gravitational waves in a spatially flat FRW background can be analyzed  by
considering a  transverse and traceless perturbation of the metric.  
Rewrite Eq.~(\ref{modifiedeinsteinequation})  and define an effective energy momentum tensor, $T^{\text{eff}}_{\mu\nu}$, by
\begin{equation}
G_{\mu\nu}=\frac{1}{M_{pl}^2}T^{\text{eff}}_{\mu\nu}\equiv \frac{a_2}{1+a_2R}\left(-\frac{1}{4}R^2g_{\mu\nu}-g_{\mu\nu}\nabla^{\alpha}\nabla_{\alpha}R+\nabla_{\mu}\nabla_{\nu}R\right)\,.
\end{equation} 
One may express $T^{\text{eff}}_{\mu\nu}$ in terms of effective fluid quantities and describe the perturbations 
of the above equation using a gauge invariant formulation. Let $H_{{\bf{k}}}({\bf{x}},t) \propto \text{exp}(i{\bf{k}}\cdot{\bf{x}}) u_{k}(t)$
be the graviton mode function. Then the evolution of $u_{k}(t)$ is given by~\cite{HawngNoh:1996, Hawng:1991}  
\begin{equation}
\frac{1}{a^3(t)F(R)}\left[ a^3(t)F(R)\partial_{t}u_k\right(t)],_{t} + \frac{k^2}{a(t)^2}u_k(t)=0\,,
\label{101Hawng}
\end{equation}
where $F(R)\equiv d f(R)/dR$. Note that this equation differs from the general relativity case by an extra term,   
 $(F,_{t}/F)(\partial_{t} u_k)$, which comes from the nonzero anisotropic pressure part of the imperfect fluid $T_{\mu\nu}^{\text{eff}}$. 
 Defining $u_k(\eta) =  v_k(\eta) /(a\sqrt{F}) $,  Eq.~(\ref{101Hawng}) becomes
\begin{equation}
\frac{d^2 v_k(\eta)}{d\eta^2} + \left[k^2 -\frac{1}{a\sqrt{F}}\frac{d^2 (a\sqrt{F})}{d\eta^2}\right]v_k(\eta)=0 \,.
\label{115Hawng}
\end{equation}
In the limit $a_2 \rightarrow 0$, where $F \rightarrow 1$, we recover the known results from general relativity. 
In this limit, after setting 
$\chi_k(\eta) = v_k(\eta)$ and using  $R(\eta)=6\ddot{a}(\eta)/a(\eta)^3$, Eq.~(\ref{115Hawng}) becomes Eq.~(\ref{xkequation}),
 as expected. 

The easiest way to analyze the behavior of oscillations in this model in a spatially flat FRW background is to take advantage of the 
equivalence between $f(R)$ theories and  scalar-tensor gravity. [For a review and discussion about $f(R)$ gravity and its equivalence 
with the scalar-tensor theory for gravitation see Refs.~\cite{DeFeliceTsujikawa:2010,Faulkneretal:2006}.] 
The usual approach to obtain a scalar-tensor gravity from $f(R)$ gravity is to perform a conformal transformation, 
$\tilde{g}_{\mu\nu}=F(R)g_{\mu\nu}$ with $F(R)\equiv df(R)/dR$, and to introduce an auxiliary scalar field $\phi$ according 
to $F(R(\phi))=\text{exp}[\sqrt{2/3}~(\phi/M_{pl})]$. In the new frame, or Einstein frame, the theory looks like conventional 
general relativity plus a minimally coupled auxiliary scalar field $\phi$. It is, however, not identical to the GRSF model of the
previous subsection. Note that we use $\varphi$ to denote the scalar field in the GRSF model, and $\phi$ to denote that in the $f(R)$ model. The scalar field $\phi$ can oscillate around the minimum of its potential, which leads to  oscillatory behavior 
of the scale factor in the original frame, or Jordan frame, of the form     
\begin{equation}
a(t) = \bar{a}(t) \left[ 1 -  E_i\, \left(\frac{ \bar{a}_i }{\bar{a}(t)}\right)^{3/2} \,   \cos(\omega t)   \right]\,,
\label{scalefactorf(R)}
\end{equation}
 where $E_i=(\phi_i)/(\sqrt{6}M_{pl})$ is the initial amplitude of metric oscillations, $\phi_i > 0$ is the initial value of the scalar field,  
 $\bar{a}(t)$ is the background scale factor time averaged over the oscillations, and $\bar{a}_i = \bar{a}(t_i)$ where $t_i$  is the time 
 at which oscillations start. 
The equivalence between $f(R)$ gravity and scalar-tensor gravity and the derivation of Eq.~(\ref{scalefactorf(R)}) are discussed in 
detail in  Appendix~\ref{f(R)gravity}.
 If we consider the background scale factor to be  flat spacetime and $E_i \ll 1$, Eq.~(\ref{scalefactorf(R)}) takes the form, to 
 leading order, of Eq.~(\ref{scalefactor})  with $E_i = A_0$ and $\omega = \omega_0$, where $\omega$ is the mass of the scalar 
 field $\phi$. Thus in the $f(R)$ model, the scale factor and the scalar field oscillate at the same frequency.
 
\begin{figure}
\centering
\includegraphics[scale=0.7]{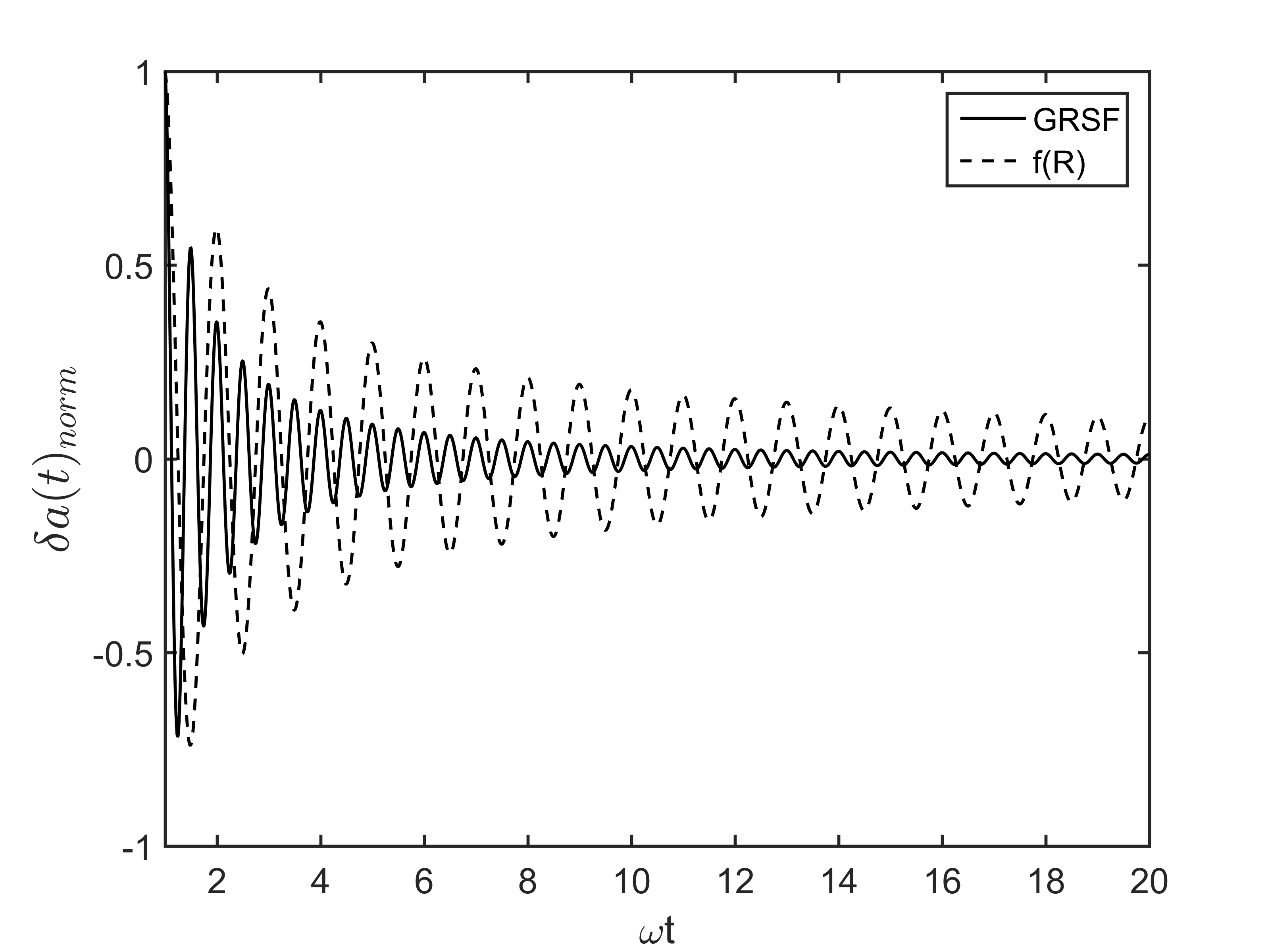}
\caption{Oscillatory part of the scale factor in a radiation dominated universe
is illustrated in the GRSF model and in $f(R)$ gravity.   Here $\delta a(t)_{\text{norm}}$
is $\delta a(t)$ expressed in units where $\delta a(t)_{\text{norm}} = 1$ at an initial time given by $\omega t =1$.}
\label{figure3}
\end{figure}

 We can express the scale factors of both models, Eqs.~(\ref{scalefactorGRscalar}) and (\ref{scalefactorf(R)}), as 
 \begin{equation}
 a(t) = \bar{a}(t)[1 + \delta a(t)]\, ,
 \label{eq:del-a}
 \end{equation}
 where $\delta a(t) \ll 1$ is the oscillatory part of the scale factor. 
 Figure~\ref{figure3} illustrates the behavior of this oscillatory part in both models  in a radiation dominated universe. 
 Here $\bar{a}(t) \propto t^{1/2}$, so $\delta a(t) \propto  \bar{a}(t)^{-3} \propto t^{-3/2}$ in the GRSF model and
 $\delta a(t) \propto  \bar{a}(t)^{-3/2} \propto t^{-3/4}$ in the  $f(R)$ gravity model. Thus the oscillations are at twice the
 frequency and decay more rapidly in the GRSF model as compared to the  $f(R)$ gravity model.
 
\subsection{Calculation of graviton creation caused by oscillations around flat spacetime}

We consider the graviton creation in both models using the oscillating 
scale factor defined by  Eq.~(\ref{scalefactor}), which refers to small oscillations around a flat spacetime. Note that even though oscillations are present in both scenarios, the gravitational wave equation, which rules the graviton creation, is different for each case. 
    
\subsubsection{Graviton creation in standard general relativity plus a minimally coupled scalar field}    
    
We analyze the asymptotic behavior of the number and energy density of created gravitons on time scales long compared
to the period of oscillation.  
From Eqs.~(\ref{vetaequation}) and ~(\ref{ricciscalar}), the expression for $V(\eta)$ is 
\begin{equation}
V(\eta)= \frac{1}{2C(\eta)^2}\left[ \ddot{C}(\eta)C(\eta)-\frac{1}{2}\, \dot{C}(\eta)^2 \right] \, .
\end{equation}
Substituting the expression for $a(\eta)$,  Eq.~(\ref{scalefactor}), into this equation, we obtain, to first order in $A_0$,  
\begin{equation}
V(\eta)= -\frac{A_0\,\omega_0^2\cos(\omega_0\eta)}{1+A_0\cos(\omega_0\eta)}\approx -A_0\,\omega_0^2\cos(\omega_0 \eta) \, .
\label{vetaapproximation} 
\end{equation}
Here we treat the case of oscillations around flat spacetime, and hence set
$a(\eta) = 1$ in the prefactors to the integrals in  Eqs.~(\ref{densitynumbercoordinatespace}) and 
(\ref{energydensitycoordinatespace}). The graviton number density becomes 
\begin{align}
n &\approx   \frac{1}{8\pi}\times \int d\eta_1 \left[ -A_0\,\omega_0^2\cos(\omega_0 \eta_1)  \right]^2 \,  \\
&= \frac{A_0^2\,\omega_0^4}{8\pi}\times \int d\eta_1 \frac{1+\cos(2\omega_0 \eta_1)}{2} \, ,
\end{align}
where the integral on $\eta_1$ is to be taken over a long, but finite interval. On time scales long compared to $\omega_0^{-1}$, the average  number density creation rate is the same in both conformal time $\eta$ and comoving time $t$ and given by
\begin{equation}
\frac{dn}{d\eta}\Bigr|_{\text{GRSF}} = \frac{dn}{dt}\Bigr|_{\text{GRSF}} = \frac{A_0^2\,\omega_0^4}{16\pi} \, .
\label{eq:number-rate}
\end{equation}
If the oscillations last for a comoving time $t$, then the number density of created gravitons becomes, to leading
order in $A_0$,
\begin{equation}
n_{g}\sim\frac{A_0^2\,\omega_0^4t}{16\pi} \, .
\label{gravitonnumberdensity}
\end{equation} 
Let $\lambda = 2\pi/\omega_0$ be the period of oscillation, and,  in $c=1$ units, the wavelength associated with angular 
frequency $\omega_0$. The number density creation rate of Eq.~(\ref{eq:number-rate}) can be expressed as
\begin{equation}
\frac{dn}{dt}\Bigr|_{\text{GRSF}}= \, \pi^3\, A_0^2\, \lambda^{-3} \, \lambda^{-1}     \, .
\label{eq:number-rate2}
\end{equation} 
This result tells us that an average of $\pi^3\, A_0^2$ gravitons are created in volume $ \lambda^{3}$ per
oscillation.  

For the case of the graviton energy density, we use the approximate expression of $V(\eta)$ in Eq.~(\ref{vetaapproximation}) 
and calculate its derivative with respect to the conformal time, 
\begin{equation}
\dot{V}(\eta)  \approx A_0\,\omega_0^3\sin(\omega_0\eta) \, .
\end{equation}
Substituting this equation into Eq.~(\ref{energydensitycoordinatespace}), we now have for the  graviton energy density 
\begin{equation}
\rho \approx -\frac{A_0^2\,\omega_0^6}{16\pi^2}\int d\eta_1 \sin(\omega_0\eta_1) \int d\eta_2 
\frac{\ln\left|( \eta_2-\eta_1 )\mu \right|^2}{2} \sin(\omega_0\eta_2) \, ,
\label{eq:rho}
\end{equation}
where the integrals on $\eta_1$ and $\eta_2$ are to be taken over long, but finite intervals. First, let us focus on the inner integral
\begin{equation}
I(\eta_1)= \int_{-T}^{T} d\eta_2 \frac{\ln\left|( \eta_2-\eta_1 )\mu \right|^2}{2} \sin(\omega_0 \eta_2) \, ,
\label{Iintegralprevious}
\end{equation}
where we examine the limit $T \rightarrow \infty$  for fixed $\eta_1$. With the change of variable $y=\omega_0 \eta_2$, we have
\begin{align}
I(\eta_1)&=\omega_0^{-1} {\rm Re}\int_{-T\omega_0}^{T\omega_0} dy \left\lbrace \frac{\ln\left[(y-\omega_0\eta_1)\right]^2}{2} +
 \ln\left(\frac{\mu}{\omega_0}\right)  \right\rbrace  \sin(y) \, \\
&=\omega_0^{-1}  {\rm Re}  \int_{-T\omega_0}^{T\omega_0} dy \frac{\ln\left[(y-\omega_0\eta_1)\right]^2}{2}  \sin(y) \,\\
&\sim -\frac{\pi}{\omega_0}\,\sin(\omega_0\eta_1) + O\left(\frac{1}{T\omega_0}\right) \, ,
\label{Iintegral}
\end{align}
where we have, in the second line, dropped the $\mu$ dependent part because it is proportional to 
$\int_{-T\omega_0}^{T\omega_0} dy \sin(y) = 0$, and, in the third line, used the asymptotic values at $\pm\infty$ of  the cosine 
and sine integral functions. Note that the assumption of $\omega_0 \gg T^{-1}$ in Eq.~(\ref{Iintegral}) makes the integrand
in the expression for the graviton energy density, Eq.~(\ref{eq:rho}), approximately local.
Now, we have 
\begin{equation}
\rho \sim \frac{A_0^2\,\omega_0^5}{16\pi} \int d\eta_1 \, \sin^2(\omega_0\eta_1) \, .
\end{equation}
Again, on long time scales,  the average energy density creation rate is the same in both conformal and comoving time, so
\begin{equation}
\frac{d\rho}{d\eta}\Bigr|_{\text{GRSF}}=\frac{d\rho}{dt}\Bigr|_{\text{GRSF}}=\frac{A_0^2\,\omega_0^5}{32\pi} \, .
\label{eq:rate}
\end{equation}
The leading order for the graviton energy density after a time $t$ is 
\begin{equation}
\rho_{g} \Bigr|_{\text{GRSF}}
\sim\frac{A_0^2\,\omega_0^5t}{32\pi} \, .
\label{oldgravitonenergydensity}
\end{equation}
Here we are ignoring any possible interference terms. That is, we assume that the energy density of gravitons created at
earlier times adds incoherently to that of gravitons created later.
 
Equations~(\ref{eq:number-rate}) and (\ref{eq:rate}) show that the graviton number density creation rate, as 
well as the energy density creation rate, are proportional to the square of the metric oscillations $A_0$,  and that the 
mean graviton energy
is $\omega_0/2$. This latter result can be explained using the analogy with the spontaneous parametric down-conversion 
in nonlinear optics, where a nonlinear crystal is used to split photon beams into pairs of photons. Here, in accordance with the 
law of conservation of energy, the sum of the energies of the pair equals the energy of the original photon. Graviton production 
in pairs with energy $\omega_0/2$ per particle has previously been found in the context of the Starobinsky model for inflation \cite{Vilenkin:1985}.   

\subsubsection{Graviton creation in $f(R)$ gravity}

Now we can obtain the number and energy density creation rate in $f(R)$ gravity from those in the GRSF model. Substituting the expression for $V(\eta)$, Eq.~(\ref{vetaapproximation}), into the gravitational wave equation in standard general relativity, Eq.~(\ref{xkequation}), we obtain for the GRSF model
\begin{equation}
\frac{d^2 v_k(\eta)}{d\eta^2} + \left[k^2 +A_0\omega_0^2\cos(\omega_0\eta)\right]v_k(\eta)=0 \,.
\label{GRcase}
\end{equation}
For the case of $f(R)$ gravity, using $F(R) = 1 + a_2R$ and working to second order in $A_0$ in the term $(a\sqrt{F}),_{\eta \eta}/(a\sqrt{F})$ in the modified gravitational wave equation, Eq.~(\ref{115Hawng}), we have
\begin{equation}
\frac{d^2 v_k(\eta)}{d\eta^2} + \left[k^2 -3A_0^2\omega_0^2\cos(2\omega_0\eta)\right]v_k(\eta)=0 \,.
\label{fRcase}
\end{equation}
The difference between Eqs.~(\ref{GRcase}) and (\ref{fRcase}) lies in their respective sinusoidal factors.
Note that the overall sign is not important and does not change the particle creation rate. 
Making the replacements $\omega_0\rightarrow 2\omega_0$ and $A_0 \rightarrow (3/4)A_0^2$ in Eqs.~(\ref{eq:number-rate})
and (\ref{eq:rate}) for the GRSF model, we can obtain the corresponding results for $f(R)$ gravity:
\begin{equation}
\frac{dn}{dt} \Bigr|_{\text{f(R)}} = \frac{9A_0^4\omega_0^4}{16\pi} \,,
\label{eq:ratef(R)casenumber}
\end{equation}
\begin{equation}
\frac{d\rho}{dt} \Bigr|_{\text{f(R)}} = \frac{9A_0^4\omega_0^5}{16\pi}\,.
\label{eq:ratef(R)case}
\end{equation}
These last equations show that the graviton number density and energy density creation rates are proportional to the
 fourth power of the metric oscillation amplitude, $A_0$, and that the mean graviton energy is $\omega_0$.

\section{Graviton energy density in an expanding universe}
\label{scalinggravitondensity}

Now we wish to extend the results for the energy density creation rate in flat spacetime obtained in both cases, Eqs.~(\ref{eq:rate}) 
and (\ref{eq:ratef(R)case}), to 
an expanding universe. The general scale factor in a spatially flat FRW background is given by Eq.~(\ref{scalefactorinanexpandinguniverse}),
where the amplitude of the oscillations decreases with time. 
So long as the expansion rate of the background is slow compared to the oscillation rate,
\begin{equation}
\frac{1}{\bar{a}(t)}\, \frac{d \bar{a} }{dt} \ll \omega \,,
\label{eq:expandrate}
\end{equation}
we may treat the background spacetime as approximately flat, and use the results of  Eqs.~(\ref{eq:rate}) and (\ref{eq:ratef(R)case}) with 
$A_0 \rightarrow A_{\text{eff}}(t)$. Recall that $A_{\text{eff}}(t) = D_i (\bar{a}_i/\bar{a})^3$ in the GRSF model and $A_{\text{eff}}(t) = E_i (\bar{a}_i/\bar{a})^{3/2}$ in the case of $f(R)$ gravity. Then the energy density creation rates in the expanding universe become
\begin{equation}
\frac{d\rho}{dt} \approx  
J\omega_0^5 \left[\frac{\bar{a}_i}{\bar{a}(t)}\right]^6 \,,
\label{eq:ratewithdampingGR}
\end{equation}
where $J = (D_i^2)/(32 \pi)$ in the GRSF model and  $J = (9E_i^4)/(16\pi)$ in $f(R)$ gravity. Note that $d\rho/dt \propto \bar{a}^{-6}$
in both cases.
  
In addition to the damping effect on the metric oscillations, the expansion causes redshifting and dilution of the created gravitons.
After creation, the graviton energy density scales as $1/\bar{a}^4(t)$.
 Including both effects, the energy density at $t=t_0$ due to gravitons created in an interval $dt$ at an earlier time $t$ is 
\begin{equation}
d\rho_g(t_0)= J\omega_0^5\, \left[  \frac{\bar{a}_i}{\bar{a}(t)}\right]^6 \left[ \frac{\bar{a}(t)}{\bar{a}_0}\right]^4 dt \,,
\label{creationrateGR}
\end{equation}
where $\bar{a}_0 = \bar{a}(t_0)$. If we take $t_0$ to be the present time, the gravitons in question were created at redshift $z$,
where $1+z = \bar{a}_0/ \bar{a}(t)$.
These expressions tell us that the present contribution of earlier graviton production is suppressed by a factor of $( 1+z)^{-4}$ due to 
 redshifting and increased by a factor proportional to $( 1+z)^{6}$ due to the greater oscillation amplitude at earlier times. 
 
If we substitute into Eq.~(\ref{creationrateGR}) the values of $\omega_0$ and $J$ for each model, which depend upon the scalar field initial values, 
either $\varphi_i$ or $\phi_i$, we find that
 the energy density creation rate in the $f(R)$ gravity case is  4 times that in the GRSF model, if the scalar field masses 
 and initial values  are the same. Specifically we have
\begin{equation}
d\rho_g(t_0)\Bigr|_{\text{GRSF}}= \frac{\varphi_i^4\,\omega^5}{256\pi M_{pl}^4}\, \left[  \frac{\bar{a}_i}{\bar{a}(t)}\right]^6 \left[ \frac{\bar{a}(t)}{\bar{a}_0}\right]^4 dt \,,
\label{creationrateGR2}
\end{equation}
\begin{equation}
d\rho_g(t_0)\Bigr|_{\text{f(R)}}= 4\times\frac{\phi_i^4\,\omega^5}{256\pi M_{pl}^4}\, \left[  \frac{\bar{a}_i}{\bar{a}(t)}\right]^6 \left[ \frac{\bar{a}(t)}{\bar{a}_0}\right]^4 dt \,.
\label{creationratef(R)2}
\end{equation} 
  
If the oscillations start at time $t_i$, then the graviton energy density at time $t_0$ will be given by
\begin{equation}
\rho_g(t_0) =  J\omega_0^5 a_i^6 \,\int_{t_i}^{t_0} \bar{a}(t)^{-2} dt \, ,
\label{scalinggravitonenergydensityGR}
\end{equation}
with $\bar{a}_0 =1$. We assume that $t_i$ is after the end of inflation and that gravitons created at earlier times do not cause 
interference  with gravitons created at later times, as was assumed in Eq.~(\ref{oldgravitonenergydensity}).

Consider a model of the universe which is spatially flat and contains radiation (photons, neutrinos, and gravitons), 
nonrelativistic matter (baryonic and nonbaryonic dark matter) and a cosmological constant associated with the dark energy. 
The model is first radiation dominated, then nonrelativistic matter dominated, and is now entering into its dark energy 
dominated phase. On time scales much longer than the period of oscillations, the Friedmann equation in this model of universe, 
which is derived in detail in the Appendices,  can be expressed as 
   
\begin{equation}
3\bar{H}(t)^2 M_{pl}^2 \approx \frac{\rho_{r,0}}{\bar{a}^4(t)} +\frac{\rho_{m,0}}{\bar{a}^3(t)}+\rho_{\Lambda,0} + \frac{\omega^2\chi_i^2}{2}\left( \frac{\bar{a}_i}{\bar{a}} \right)^3\, ,
\label{Friedmannequation} 
\end{equation}
where $\bar{H}(t)\equiv[\dot{\bar{a}}(t)]/[\bar{a}(t)]$ is the Hubble parameter as a function of the time-averaged scale factor,
$\bar{a}(t)$. Here  $\rho_{r,0}$,   $\rho_{m,0}$,
and $\rho_{\Lambda,0}$ are the  radiation, nonrelativistic matter, and dark energy densities today,  respectively, and the
scalar field energy density is $\rho_{\chi} \approx (\omega^2 \chi_i^2/2) (\bar{a}_i/\bar{a})^3$, where $\chi$ refers to either the 
$\varphi$ scalar field in the GRSF model or the $\phi$ scalar field in $f(R)$ gravity. 

Since we are interested in cosmological implications of the quantum graviton creation,  we assume that oscillations of the scale 
factor continue through the present epoch. This is equivalent to requiring 
 that the scalar field in each model continues in its oscillatory phase. Note that the scalar energy density in both cases scales 
 like nonrelativistic matter, and could grow to dominate the radiation energy density before the expected beginning 
 of the matter-dominated epoch. 
 In order to avoid that, the scalar energy density, $\rho_{\chi}(t)$, should be always less than that of the  nonrelativistic matter, 
 $\rho_m(t)$,  through the present epoch. Indeed, this conclusion is supported by observational data, as will be explained in detail 
 in Sec.~\ref{cosmlogicalconstraints}. 

If we assume $\rho_{\chi} (t)  < \rho_{m}(t)$, the Friedman equation for both models, Eq.~(\ref{Friedmannequation}), becomes
\begin{equation}
\frac{\bar{H}(t)^2}{H_0^2} \approx \frac{\Omega_{r,0}}{\bar{a}^4(t)} +\frac{\Omega_{m,0}}{\bar{a}^3(t)}+\Omega_{\Lambda,0} \, ,
\label{Friedmannequationapproximation} 
\end{equation}
where  $\Omega_{r,0}=\rho_{r,0}/\rho_{c,0}$, $\Omega_{m,0}=\rho_{m,0}/\rho_{c,0}$, and 
 $\Omega_{\Lambda,0}=\rho_{\Lambda,0}/\rho_{c,0}$. Here $\rho_{c,0} = (3  H_0^2)/(8 \pi G)$ is the critical density today 
 and $G$ is  Newton's constant. Then 
 $\Omega_0=\Omega_{r,0}+\Omega_{m,0}+\Omega_{\Lambda,0} \approx 1$ is the energy density parameter today.
We use the values $H_0\equiv 100~h_0\text{ km}\,\text{s}^{-1}\,\text{Mpc}^{-1}$, $\Omega_{r,0} =4.15\times10^{-5}~h_0^{-2}$, 
 and $\rho_{c,0}=1.88\times10^{-26}~h_0^2\text{ kg m}^{-3}$. We take $h_0 = 0.673$ and     $\Omega_{m,0} = 0.315$ from the 
Planck temperature power spectrum data including WMAP polarization at low multipoles~\cite{Planckdata:2013}.  
 
Substituting Eq.~(\ref{Friedmannequationapproximation}) into Eq.~(\ref{scalinggravitonenergydensityGR}), the graviton energy density today is  found
 to be
\begin{equation}
\rho_g(t_0) = \frac{J\omega_0^5 \bar{a}_i^6}{H_0}\, \int_{\bar{a}_i}^{1} \left(  \frac{\bar{a}^{-1}}{\sqrt{{\Omega_{r,0}}+
{\Omega_{m,0}}\,{\bar{a}}+\Omega_{\Lambda,0}\, \bar{a}^4}}\right) d\bar{a} \, .
\label{rho_g(t_0)G(R)} 
\end{equation}
 This integral cannot be expressed in terms of elementary functions and must be calculated numerically.
The graviton energy density during the radiation dominated epoch can be calculated more easily. At some time $t_{r} \alt t_{rm}$, 
where $t_{rm}$ is the  time of radiation-matter equality, the scale factor can be approximated as 
 \begin{equation}
\bar{a}(t)\approx{({2\sqrt{\Omega_{r,0}}\, H_0\, t})^{1/2}}  \propto   \sqrt{t} \,.
\label{eq:root-t}
\end{equation}
 This is a solution of Eq.~(\ref{Friedmannequationapproximation}) when the nonrelativistic matter and dark energy terms may be neglected 
 compared to the radiation term, and the latter term is assumed
to come entirely from photons and neutrinos. If other relativistic particles are present, then the constant of proportionality increases
by a factor of the fourth root of the number of types of particles present. This factor will be assumed to be of order
one, and will be ignored in our rough estimates.

As a result, the graviton energy density at time $t_r \gg t_i$ is given by
\begin{equation}
\rho_g(t_r) = J\omega_0^5\int^{tr}_{t_i}\left[ \frac{\bar{a}_i}{\bar{a}(t)} \right]^6 \left[ \frac{\bar{a}(t)}{\bar{a}(t_r)} \right]^4dt \approx 
J\omega_0^5\left(  \frac{t_i^{3}}{t_r^{2}}\right)\ln({t_r/t_i})\,.
\label{rho_g(t^*)GR}
\end{equation}
Here we are assuming that the oscillations begin during the radiation dominated era. Clearly some significant event is needed to cause the oscillations
to begin and to determine the initial amplitude. Two possibilities are the reheating at the end of inflation, or a subsequent
phase transition. Note that the graviton energy density in Eq.~(\ref{rho_g(t^*)GR}) vanishes in the limit $t_i \rightarrow t_r$ as is expected. 

Thus far we have not discussed the decay of the scalar fields caused by direct coupling with other fields such as radiation or 
nonrelativistic matter and/or the quantum particle production different from gravitons. Even though  in the GRSF model we have 
not considered a direct coupling between the scalar field and matter fields, the field $\varphi$ couples with those fields through 
gravity by means of the scale factor (the oscillatory part of the scale factor is proportional to $\overline{\varphi^2}$). This coupling 
results in quantum particle production not only of gravitons (when the scale factor coupling to a pair of minimally coupled massless
 scalar fields) but also, for instance, of massive scalar particles, vector bosons and fermions~\cite{Ema:2015}. In any case, if we 
 are interested in values for $\omega$ below the masses of these particles, we expect that these processes are 
 mass suppressed. We have a similar scenario for $f(R)$ gravity, with the difference that in this theory there is a direct coupling
 between the auxiliary scalar field $\phi$ and the matter fields.
However this coupling is suppressed in the regime in which we are working, where $E_i \propto (\phi_i/M_{pl}) \ll  1$. 

\section{Cosmological constraints on the oscillation amplitude of the scale factor}
\label{cosmlogicalconstraints}

In this section, we  explore three  cosmological constraints on the graviton creation. The first two are observational
constraints on the effects of the created gravitons, 
one from big bang nucleosynthesis (BBN) and another from observational Hubble parameter measurements. The third
comes from an observational constraint on scalar field energy density, which in the context of the specific models we
treat, implies a strong constraint on the amplitude of oscillations.
 All of these constraints will depend on the value for $\omega$ considered. 
In $f(R)$ gravity the angular frequency of oscillations is bounded from below, $\omega \geq  \omega_B$. There is no analogous
bound in  the GRSF model, but in both models we will consider a range of angular frequencies beginning at $\omega_B$
and extending upward by several orders of magnitude. The upper bound on  $\omega$ could be as high as the Planck
frequency, $10^{31}\,  \omega_B$, where our semiclassical approach is expected to break down. However, we will be
primarily concerned with more typical particle physics energy scales.

\subsection{Big bang nucleosynthesis constraint}

Commonly, the BBN bound is expressed as a number of extra neutrino varieties, 
$\Delta N_{\nu}$. (For a review, see big bang cosmology and big bang nucleosynthesis reviews in Ref.~\cite{Oliveetal:2014}.)
 In the early universe, relativistic particles dominate the total energy density. For this reason, at $T=1$ MeV (before 
 electron-positron annihilation), the total energy density   is   $\rho_{BBN} = N(T)(\pi^2/30) T^4$, where $N(T)$ is the 
 equivalent number of degrees of freedom at temperature $T$,  approximately given by the contribution of photons, electrons, 
 positrons and neutrinos. Any additional contribution at that time to the total energy density from a component with a radiation-like
 equation of state can be described as an equivalent number of extra neutrinos. Thus, the graviton energy density $\rho_{gBBN}$ 
 at $T=1$ MeV is
\begin{equation}
\rho_{gBBN} = \frac{7}{8}\Delta N_{\nu} \, \rho_{\gamma} \, , 
\label{gravitonenergydensityBBN}
\end{equation}
where $\rho_{\gamma}=\left[(2\pi^2)/(30)\right]T^4$ refers to the photon energy density.

It is possible to find in the literature several constraints on $\Delta N_{\nu}$, which depend upon the specific light element
abundances considered,  from  $\Delta N_{\nu} \leq 0.2$ to $\Delta N_{\nu} \leq 1$~\cite{Giovannini:2010}. The constraint 
 can be relaxed in some nonstandard nucleosynthesis scenarios~\cite{Giovannini:2002}.  We take for our purpose 
 $\Delta N_{\nu} \approx 1$. Then, using Eq.~(\ref{rho_g(t^*)GR})  for the graviton energy density in the radiation-dominated 
 epoch, we have
\begin{equation}
\rho_g(t_r) \approx J\omega_0^5 \left(\frac{t_i^{3}}{t_r^{2}}\right)\ln{(t_r/t_i)} \leq \frac{7}{8} \rho_{\gamma} \, ,\\
\label{gravitonenergydensityBBNboundGR}
\end{equation}
where $t_r$ refers to the time when $T=1$ MeV, which is approximately one second.
  Equation~(\ref{gravitonenergydensityBBNboundGR})  gives a bound on $D_i$, in the GRSF model, and $E_i$, in $f(R)$ gravity, for a given $\omega$ of
\begin{equation}
D_i\Bigr|_{\text{G(R)}}\, \alt 10^{-5}\, 
 \left(\frac{10^{-6} \,{\rm s}}{t_i}\right)^\frac{3}{2}\, \left(\frac{10^{10}\, \omega_B}{\omega}\right)^\frac{5}{2}\,
 \left[\ln{(1~ \text{s}/t_i)}\right]^{-1/2} , 
 \label{eq:BBN-boundGR}
\end{equation}
\begin{equation}
E_i\Bigr|_{\text{f(R)}}\, \alt 3\times 10^{-3}\, 
 \left(\frac{10^{-6} \,{\rm s}}{t_i}\right)^\frac{3}{4}\, \left(\frac{ 10^{10}\, \omega_B}{\omega}\right)^\frac{5}{4}\,
 \left[\ln{(1~ \text{s}/t_i)}\right]^{-1/4}  \, . 
 \label{eq:BBN-boundfR}
\end{equation}
Recall that the initial oscillation amplitude, $D_i$ and $E_i$, needs to be small for the consistency of our treatment. This
condition can be fulfilled if $\omega \agt 10^{10}\, \omega_B \approx 26\, {\rm MeV}$.
Note that an initial time $t_i = 10^{-6} \,{\rm s}$ corresponds to a temperature of $T_i \approx 1\, {\rm GeV}$. 

\subsection{Constraint from the expansion rate of the universe}

Observational data on the late universe can be used to obtain an upper bound on the present density of gravitons. 
Rewriting the scale factor as a function of the redshift in Eq.~(\ref{Friedmannequationapproximation}) using 
$\bar{a}(z)=1/(1+z)$, we obtain 
\begin{equation}
\bar{H}(z)=H_0\left[\Omega_{r,0}(1+z)^4 +\Omega_{m,0}(1+z)^3+ (1-\Omega_{r,0}-\Omega_{m,0})\right]^{1/2} \, ,
\label{H(z)} 
\end{equation}
which shows the dependence of $\bar{H}(z)$ on the cosmological parameters. Taking into account graviton production,
 Eq.~(\ref{H(z)}) becomes
\begin{equation}
\bar{H}(z)=H_0\left[(\Omega_{r,0}+\Omega_{g,0})(1+z)^4 +\Omega_{m,0}(1+z)^3+ (1-\Omega_{r,0}-\Omega_{m,0}-\Omega_{g,0})\right]^{1/2} \, ,
\label{H(z)plusgraviton} 
\end{equation}
where $\Omega_{g,0}$ is the graviton energy density parameter today. 

We use a sample of 18 observational measurements of Hubble parameter in the range of $0.09 \leq z \leq 1.75$ with their respective
 standard errors reported by Moresco {\it et al.}~\cite{Morescoetal:2012},  Table 1. Measurements are provided 
 by passively evolving galaxies, high-quality spectra of red-envelope galaxies in galaxy clusters, and spectroscopic evolution of early
 type galaxies. The least-squares method is applied by means of minimizing the  reduced sum of the square of residuals weighted by errors $\chi^2_\nu$ according to             
\begin{equation}
\chi^2_{\nu}(\Omega_{g,0})= \frac{1}{\nu}\sum_{i=1}^{18}\frac{[H^{obs}(z_i) - \bar{H}(z_i;\Omega_{g,0})]^2}{\sigma_{H^{obs}(z_i)}^2} \, ,
\label{chisquare} 
\end{equation}
where $H^{obs}(z_i)$ is the $i$th observational value of $H(z)$ at redshift $z_i$, 
$\bar{H}(z_i;\Omega_{g,0})$ is the theoretical $i$th value of $H(z)$ obtained by means of Eq.~(\ref{H(z)plusgraviton}) at redshift $z_i$, 
$\sigma_{H^{obs}(z_i)}$ is the error associated
with the $i$th observational value of $H(z)$ at redshift $z_i$, and $\nu$ is the number of degrees of freedom (18 observational data 
points minus one parameter to be adjusted, i.e., $\Omega_{g,0}$). The standard errors, $\sigma_{\Omega_{g,0}}(\Omega^*_{g,0})$ and $\sigma_{H}(z;\Omega^*_{g,0})$, associated with the graviton energy density today and the fitted function $\bar{H}(z)$, respectively, are calculated 
following standard procedures~\cite{Richter:1995}. Here we have defined $\Omega^*_{g,0}$ as the value of the graviton energy density parameter today which minimizes $\chi^2_\nu$. 

\begin{figure}
\centering
\includegraphics[scale=0.7]{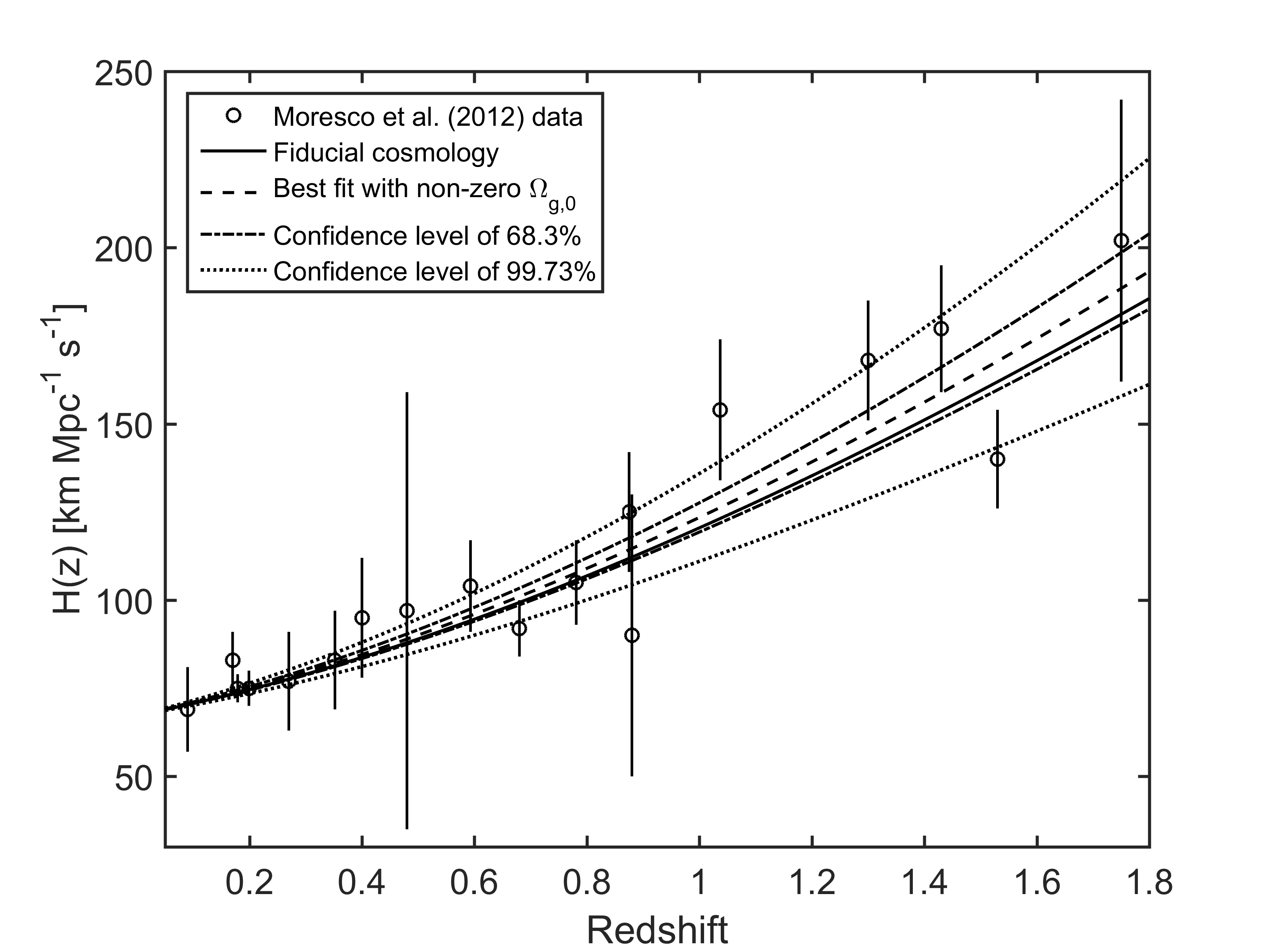}
\caption{Observational data for $H(z)$ and their errors (see Ref.~\cite{Morescoetal:2012}) are plotted. The solid line gives
 the fiducial cosmology, which assumes no gravitons with $h_0 = 0.673$, $\Omega_{r,0}=4.15\times10^{-5}~h_0^{-2}$,
  $\Omega_{m,0}=0.315$ , and $\Omega_{\Lambda}=1-(\Omega_{r,0}+\Omega_{m,0})$. The dashed line is a best fit using 
  least squares method with nonzero $\Omega_{g,0}$. The boundaries of the regions associated with confidence levels 
  of $\sigma_{H}(z;\Omega^*_{g,0})$ and of  $3 \sigma_{H}(z;\Omega^*_{g,0})$  are also illustrated. }
\label{figure1}
\end{figure}

Figure~\ref{figure1} shows the fiducial cosmology without gravitons, obtained from Eq.~(\ref{H(z)}), and the best fit with a
 nonzero value for the graviton energy density parameter today, obtained from Eq.~(\ref{H(z)plusgraviton}). Including 
gravitons in the evolution of the Hubble parameter is equivalent to increasing the radiation energy density parameter. 
This produces 
an increase of the Hubble parameter for a given $z$  in comparison to the fiducial cosmology. 
The best fit is found to be $\Omega^*_{g,0} = 0.011 \pm 0.015$ (for 1 standard deviation) with $\chi^2_\nu = 0.75$. 
The value of $\chi^2_\nu$ is reasonably close to 1 indicating that the fit can be considered meaningful. (See, for example,
Ref.~\cite{Richter:1995}.)    
At the level of two standard deviations, we obtain an upper bound for the graviton energy density parameter today of 
$\Omega^*_{g,0} \leq 0.04$. 

Because the graviton energy density increases as the comoving time increases, it is in principle possible to obtain constraints 
on the oscillation amplitude for each case:
\begin{equation}
\frac{\rho_{g}(t_0)}{\rho_{c,0}} = \Omega^*_{g,0} \lesssim 0.04 \, .
\label{newconstraint} 
\end{equation}
 Use Eq.~(\ref{rho_g(t_0)G(R)}) for $\rho_{g}(t_0)$. Then the constraints on the oscillation amplitudes may be expressed as
\begin{equation}
D_i\Bigr|_{\text{GRSF}} \alt  10^{-5} \,  
 \left(\frac{T_i}{1 \, {\rm GeV}}\right)^3   \, \left(\frac{ 10^{10}\, \omega_B}{\omega}\right)^\frac{5}{2}\,,
 \label{constraintlate1}
\end{equation}
\begin{equation}  
E_i\Bigr|_{\text{f(R)}} \alt  10^{-2}\,
 \left(\frac{T_i}{1 \, {\rm GeV}}\right)^\frac{3}{2}   \,\left(\frac{10^{10}\,  \omega_B}{\omega}\right)^\frac{5}{4}\,  .
\label{constraintlate2}
\end{equation}
Here we have used $\bar{a}_i \approx 3K/T_i$, where $T_i$ is the initial energy scale, for the factor $\bar{a}_i^6$ in 
Eq.~(\ref{rho_g(t_0)G(R)}). Moreover,  since the definite integral in this equation
is slowly varying with respect to its lower limit, 
$\bar{a}_i$, we have evaluated it at  $T_i$ =  1 GeV, where its value is about $2 \times 10^3$.

Note that these constraints from late time dynamics of the universe are comparable to those obtained from nucleosynthesis
in Eqs.~(\ref{eq:BBN-boundGR}) and  (\ref{eq:BBN-boundfR}).   There seem to be competing effects which nearly cancel
one another. Nucleosynthesis occurs earlier in the history of the universe when then characteristic amplitude of the
oscillations is greater and there has been less redshifting of the  created gravitons. However, in the late universe, there has 
been far more time for graviton creation.

\subsection{Constraints on the scalar field energy density}

Now we consider a constraint on the scalar energy density, $ \rho_{\chi}$, and its implications. Data from the dynamics of galaxy 
clusters~\cite{Bahcalletal:2014} lead to an estimate of the current matter density of  $\Omega_{m,0} = 0.26$. This estimate
includes all matter, including dark matter, which is localized on the scale of a cluster of galaxies, but would not include
a homogeneous background density, such as that due to a scalar field. CMB data from the Planck collaboration 
2013~\cite{Planckdata:2013} leads to a slightly larger value of $\Omega_{m,0} = 0.315$. Given that about 70$\%$ of the 
current energy density is dark energy, the scalar field energy density must be less than the matter density,
\begin{equation}
\rho_{\chi}(t) < \rho_m (t) \,.
\label{eq:scalar-density}
\end{equation}
Note that this is also a  constraint on $\chi_i$, the initial value of the scalar field. Because  
$\rho_{m} \approx \rho_{m,0}/\bar{a}^3 \approx \rho_{m,0}(T/T_0)^3$ and 
$\rho_{\chi} \approx (\omega^2\chi_i^2/2)(\bar{a}_i/\bar{a})^3 \approx (\omega^2\chi_i^2/2)(T/T_i)^3 $, we have
\begin{equation}
\frac{\chi_i}{M_{pl}} \lesssim 10^{-11}~\left(\frac{T_i}{1~\text{GeV}}\right)^{3/2}\left(\frac{\omega_B} {\omega} \right)\,,
\label{firstassumption}
\end{equation}
where $T_i$ and $T_0$ are  the temperature at time $t_i$ and the current temperature of the cosmic microwave background, 
respectively. 

 This constraint on the scalar energy density leads to a very strong constraint on the initial amplitude of 
 oscillations in both models:
\begin{equation}
D_i\Bigr|_{\text{GRSF}} \alt  10^{-23} \,  
 \left(\frac{T_i}{1 \, {\rm GeV}}\right)^3   \, \left(\frac{ \omega_B}{\omega}\right)^2\,,
 \label{strongboundDi}
\end{equation}
\begin{equation}  
E_i\Bigr|_{\text{f(R)}} \alt  10^{-12}\,
 \left(\frac{T_i}{1 \, {\rm GeV}}\right)^\frac{3}{2}   \,\left(\frac{ \omega_B}{\omega}\right)\,  .
\label{strongboundEi}
\end{equation}

These constraints are much stronger than the constraints which come directly from the observable effects of the created
gravitons. This is presumably related to the weakness of the graviton creation process. However, the scalar field
 energy density constraint is more model dependent, and comes from the key role played by scalar fields in both of
 the specific models treated here.
 
\section{Quantum decoherence induced by the graviton energy density}
\label{quantumdecoherenceinducedbygravitonbath}

A realistic quantum system cannot be considered isolated, but is in  interaction with the surrounding environment. This 
interaction can induce in the system a loss of quantum coherence, namely, a local suppression of interference between 
two different states~\cite{Schlosshauer:2004}. The environment can refer to ordinary matter, quantum fields, or gravitational 
fields. For a recent review and discussion about quantum decoherence and gravitational interactions,
 see Ref.~\cite{AnastopoulosandHu:2013}.
De Lorenci and Ford~\cite{LorenciandFord:2015} studied the decoherence rate of quantum systems induced by a bath of long wavelength gravitons. The basic mechanism arises from quantum geometry fluctuations produced by the graviton bath, which
in turn produce length and hence phase fluctuations in a quantum system. These phase fluctuations lead to a loss of contrast
in interference patterns, and hence decoherence by dephasing. 
We will apply these results to quantum systems in a bath of graviton created by the mechanism discussed in the GRSF model. 
First, we summarize the essential results of Ref.~\cite{LorenciandFord:2015}. 

Adopt the transverse-tracefree gauge and 
 define $h$ as the root-mean-square fractional length fluctuations in a particular direction, such as the $x$-direction by
\begin{equation}  
h^2=\langle (h_{xx})^2 \rangle = (1/9)\langle h^{TT}_{ij}h^{ij}_{TT}\rangle \,.
\end{equation}
We can reexpress $h$ as a  function of the graviton energy density as 
\begin{equation}
h = \frac{4}{3}\sqrt{2\pi}\frac{\lambda_g \sqrt{\rho_g}}{E_p} \, ,
\label{expressionforh}
\end{equation}
where $\lambda_g = 2\pi/\omega_g$ is the characteristic graviton wavelength and $E_p$ is the Planck energy.
 Suppose we have a quantum system in which $\Delta \omega$ is the energy difference between the interfering states. 
 The decoherence time $t_d$ induced by length fluctuations  is approximately $t_d \approx 1/(h \Delta \omega)$.
If the graviton wavelength is large compared to the geometric size of the quantum system,
 the decoherence time may be written as  
\begin{equation}
t_d = \frac{3}{4\sqrt{2\pi}}\frac{E_p}{\lambda_g \sqrt{\rho_g}\Delta\omega}\, .
\label{generalformuladecoherencetime}
\end{equation} 
Note that decoherence by the effects of a graviton bath seems to be compatible with the assumption, stated after Eq.~(\ref{oldgravitonenergydensity}), that the graviton energy density accumulates incoherently. A thermal bath of gravitons is maximally incoherent, but is expected to produce length and hence phase fluctuations. The key issue is that the typical graviton wavelength be larger than the size of the quantum system.

In our case, the graviton energy density may be taken to be the present value given by Eq.~(\ref{rho_g(t_0)G(R)}), and  $\lambda_g$
is understood to be an average wavelength at the present time. For the purpose of an estimate, we take the energy density
to be at the upper bound of $4\%$ of the total energy density of the universe found in Eq.~(\ref{newconstraint}). We also take
$\lambda_g = 2\pi/\omega_g \approx 4\pi/\omega_0$. That is, we use the GRSF model, where
 the gravitons are created  with an angular
frequency of $\omega_0/2$, and we are assuming that the present graviton bath is composed of gravitons which have not been
significantly redshifted since their creation. This is reasonable, given that in the time that a given graviton's energy has been
redshifted by a factor of $1/2$, its contribution to the energy density has decreased by a factor of $1/16$. 
 With these assumptions, we obtain a lower bound on the decoherence time of
\begin{equation}
t_d \agt  10^{7} \text{ yr}\,  \left(\frac{\omega}{\omega_B}\right) \left( \frac{1 \text{ eV}}{\Delta \omega} \right)\, , 
\label{eq:td-bound}
\end{equation} 
where we have associated the mass of the scalar field $\varphi$ with the angular frequency of oscillations using  
$\omega = \omega_0/2$.
For $\omega \approx \omega_B$, this lower bound holds for quantum systems with a geometric size small compared to $\lambda_{g} \approx 0.05 \text{ cm}$. 
This decoherence time is quite long unless the energy difference $\Delta \omega$ is large.      
  
\section{Summary and discussion}
\label{summarydiscussion}

We have studied quantum creation of gravitons by small  scale factor oscillations in a spatially flat FRW background. 
We use the perturbative method
of Birrell and Davies~\cite{BD80,BirrellDavies:1982}, which is an expansion in powers of a parameter describing the deviation
from conformal coupling. In our case,
the effective expansion parameter has the value $1/6$, which should be small enough for order of magnitude estimates, but
not for precise results. 

Sinusoidal scale factor oscillations can arise in various cosmological models and we consider two examples. The first consists of
 the standard matter fields in general relativity plus the addition of a minimally coupled scalar field, $\varphi(x)$, in a harmonic
  potential (GRSF model). The second model involves a modification of Einstein gravity in which a term proportional to the
square of  the Ricci scalar is added to the gravitational action [$f(R)$ gravity model]. 
The same modified Einstein equation also arises, perhaps more naturally, in semiclassical gravity theory, where the classical
gravitational field is coupled to the renormalized expectation value of a quantum matter stress tensor. The $f(R)$ gravity model
is equivalent to a scalar-tensor theory of gravity, and the scale factor oscillations may be described in terms of oscillations
of the scalar field in the scalar-tensor theory. Laboratory tests of the
inverse square law for gravity give an upper bound on the coefficient of the $R^2$ term in $f(R)$ gravity, which leads to a 
lower bound, $\omega_B$, on the oscillation frequency $\omega$. By contrast, in the GRSF model the value of $\omega$ is 
not bounded from below. In both models the amplitude of oscillations is a free
parameter and presumably determined by  initial conditions. In the GRSF model, the quantum graviton production is ruled by the
 standard gravitational wave equation from general relativity, but in $f(R)$ gravity, the graviton creation is ruled by a modification of 
 this equation. This leads to different expressions for the graviton creation rates in the two models. In both models, the amplitude 
 of the scale factor oscillations decays as the universe expands. If $\bar{a}(t)$ is the background scale factor, time averaged
  over oscillations, then the amplitude decreases as $\bar{a}(t)^{-3}$ in the GRSF model, and as  $\bar{a}(t)^{-3/2}$ in the $f(R)$ 
 model.

We first obtained expressions for the number and energy
density creation rates on an average background of flat spacetime in both models, Eqs.~(\ref{eq:number-rate}), 
(\ref{eq:rate}), (\ref{eq:ratef(R)casenumber}), and (\ref{eq:ratef(R)case}). We then extended our analysis to an expanding universe 
by including two effects: damping of the metric oscillations and density dilution and redshifting of the created gravitons. 
The results show the differences between the two models with respect to the dependence upon initial amplitude, angular frequency, and damping rate 
of the oscillations. If the mass of the scalar field in each model is $\omega$, the angular frequency of the metric oscillations is 
2$\omega$  in the GRSF model, and $\omega$ in  $f(R)$ gravity.  The angular frequency of the created gravitons is 
$\omega$ in both models. The initial amplitude of oscillations is expected to be 
determined by processes in the early universe, such as at reheating or a subsequent phase transition. 

We assumed the matter fields in both models to be the usual perfect fluids associated with radiation,  nonrelativistic matter and 
a cosmological constant. We examined two cosmological constraints on the energy density of the created gravitons, and hence 
on the initial amplitude of the oscillations for fixed $\omega$. The first constraint comes from big bang nucleosynthesis and the second
from data on the expansion rate of the late universe. Both constraints lead to similar bounds on the initial metric oscillation
amplitudes. These bounds become meaningful if $\omega \agt 26 \,{\rm MeV}$. The expansion rate data indicate that gravitons
cannot comprise more than about $4\%$ of the present mass density of the universe. We also used data from the dynamics
of galaxy clusters and the cosmic microwave background to argue that the energy density of the scalar fields, which appear in both
of our models, must be small compared to the current density of nonrelativistic matter. This in turn places strong constraints on
the amplitudes of the scalar field oscillations, and hence on the amplitudes of the scale factor oscillations. The latter constraints
are much stronger than those obtained from the effects of the created gravitons, but are more dependent upon the details of
our specific models, and potentially less robust. 

Finally, we examined the role of the bath of gravitons produced by the GRSF model in decohering quantum systems, 
using the results of 
Ref.~\cite{LorenciandFord:2015}. Long wavelength gravitons produce quantum spacetime geometry fluctuations
which in turn lead to length and phase fluctuations in a system exhibiting quantum interference. The phase fluctuations
lead to a loss of contrast in the interference pattern. Using our upper bound on the present graviton energy density from data of the Hubble parameter in the late universe, 
 leads to a lower bound on the characteristic decoherence time, $t_d$, given in Eq.~(\ref{eq:td-bound}). 
This bound allows the decoherence time to be quite long unless the energy difference of interfering components 
of the system is large.  

\acknowledgments

We thank Mark Hertzberg, Alexander Vilenkin, and Xiaozhe Hu for valuable discussions. This work was supported in part  by the National Science Foundation under Grants No. PHY-1506066 and No. PHY-1607118.

\appendix
\section{DYNAMICS OF THE GRSF MODEL}
\label{GRscalar}
In this Appendix, we will derive Eqs.~(\ref{scalefactorGRscalar}) and (\ref{Friedmannequation}) for the GRSF model
in a spatially flat FRW background.  

Consider the Friedmann equation for this model, Eq.~(\ref{scalar1}), with the energy density for matter fields, $\rho_M$, consisting of the usual energy density components of radiation ($\rho_r = \rho_{r,0}/a^4$),      nonrelativistic matter ($\rho_m = \rho_{m,0}/a^3$), and vacuum ($\rho_{\Lambda}$). Here $\rho_{r,0}$ and
$\rho_{m,0}$ are the current energy density of radiation and nonrelativistic matter, respectively. Taking into account that $\rho_{\varphi}=(\partial_t \varphi)^2/2 + (\omega \varphi)^2/2$ and Eq.~(\ref{timeevolutionscalar}), in the regime $H \ll \omega$, we have
\begin{equation}
3H^2M_{pl}^2 \approx \bar{\rho}_{\text{total}} + \frac{3}{2}\varphi_i^2 \bar{H}\omega \left(\frac{\bar{a}_i}{\bar{a}}\right)^3\cos(\omega t)\sin(\omega t)\,,
\label{FEGRSF}
\end{equation}
where $\bar{\rho}_{\text{total}} \equiv \rho_{r,0}/\bar{a}^4 + \rho_{m,0}/\bar{a}^3 + \rho_{\Lambda,0} + (\omega^2\varphi_i^2/2)(\bar{a}_i/\bar{a})^3$ and $\bar{H} \equiv \dot{\bar{a}}/\bar{a}$. Define $a = \bar{a}\left( 1 + \delta a \right)$ where $\delta a \ll 1$ is the oscillating part of the scale factor, which leads to $H = \dot{a}/a \approx \dot{\bar{a}}/\bar{a} + \dot{\delta a} = \bar{H} + \dot{\delta a}$. After a binomial expansion in powers of $\varphi_i/M_{pl}$, Eq.~(\ref{FEGRSF}) becomes
\begin{equation}
H \approx \left(\frac{\bar{\rho}_{\text{total}}}{3M_{pl}^2}\right)^{1/2} +  \frac{\varphi_i^2~ \omega}{8 M_{pl}^2}\left(\frac{\bar{a}_i}{\bar{a}}\right)^3 \sin(2\omega t)\,.
\label{appendixscalar2}
\end{equation}
By inspection, we have that $\dot{\delta a}$ corresponds to the second term on the right side of Eq.~(\ref{appendixscalar2}). If we integrate $\dot{\delta a}$ during a period of time $\bigtriangleup t_{osc}$ greater than $1/\omega$ but much less than $1/H$, the background scale factor is essentially constant in comparison to the oscillating function $\sin(2\omega t)$. Then, setting the integration constant to be zero, we have
 \begin{equation}
 \delta a = -\frac{\varphi_i^2}{16 M_{pl}^2}\left(\frac{\bar{a}_i}{\bar{a}}\right)^3 \cos(2\omega t)\,. 
 \end{equation}    
Taking into account that $a = \bar{a}(1 + \delta a)$, we obtain the result shown by Eq.~(\ref{scalefactorGRscalar}).

On time scales much longer than $\bigtriangleup t_{osc}$, we may consider the cosmological evolution
of the model to be time averaged over oscillations and the original Friedmann equation, Eq.~(\ref{scalar1}), becomes 
\begin{equation}
3\bar{H}^2M_{pl}^2 = \frac{\rho_{r,0}}{\bar{a}^4} + \frac{\rho_{m,0}}{\bar{a}^3} + \rho_{\Lambda,0} + \frac{\varphi_i^2 \omega^2}{2}\left(\frac{\bar{a}_i}{\bar{a}}  \right)^3\,.
\label{friedmanequationscalarappendix}
\end{equation}
Note that this equation, in a rough approximation, becomes Eq.~(\ref{Friedmannequationapproximation})
when $\rho_{\varphi} (t)  < \rho_{m}(t)$.
 
\section{DYNAMICS OF THE $f(R)$ GRAVITY MODEL}  
\label{f(R)gravity}

 In this Appendix, we will derive Eqs.~(\ref{scalefactorf(R)}) and  (\ref{Friedmannequation}) for the $f(R)$ gravity model
in a spatially flat FRW background.  

The action for $f(R)$ gravity can be expressed in the Jordan frame (JF) as
\begin{equation}
 S = \frac{M_{pl}^2}{2}\int d^4x \sqrt{-g} f(R) + \int d^4x \mathscr{L}_M(g_{\mu\nu},\Psi_M)\,,
\label{f(R)action}
\end{equation}
where $\mathscr{L}_M$ is the matter Lagrangian, $\Psi_M$ are matter fields  and we set  $f(R) = R + a_2R^2/2$. 
Recall that the reduced Planck mass is $M_{pl}\equiv (8\pi G)^{-1/2}$.
Let us rewrite this action as 
\begin{equation}
 S = \int d^4x \sqrt{-g} \left[\frac{M_{pl}^2}{2}F(R) R - U(R)\right] + \int d^4x \mathscr{L}_M(g_{\mu\nu},\Psi_M)\,,
\label{f(R)arrangeaction}
\end{equation}
where $U(R)= M_{pl}^2 \left[F(R) R - f(R) \right] / 2$ with $F(R)\equiv d f(R)/d R$.
The action can be transformed to the Einstein frame (EF) by introducing the conformal transformation 
$\tilde{g}_{\mu\nu}= F(R)g_{\mu\nu}$, where $F(R)$ is the conformal factor and the tilde refers to any quantity in
 the Einstein frame. We introduce an auxiliary  scalar field $\phi$ such that 
\begin{equation}
F(R(\phi)) = {\rm e}^{\frac{\sqrt{2/3}}{M_{pl} }\, \phi} \,.
\label{auxiliarscalarfield}
\end{equation}
After some manipulation, the action of Eq.~(\ref{f(R)arrangeaction}) under the conformal transformation becomes~\cite{DeFeliceTsujikawa:2010}
\begin{equation}
S_{EF} = \int d^4x \sqrt{-g} \left[\frac{M_{pl}^2}{2}\tilde{R}-\frac{1}{2}\tilde{g}^{\mu\nu}\partial_{\mu}\phi\partial_{\nu}\phi - V(\phi)\right] + \int d^4x \mathscr{L}_M(e^{-\frac{\sqrt{2/3}\phi}{M_{pl}}}\tilde{g}_{\mu\nu},\Psi_M)\,,
\label{tensorscalaraction}
\end{equation}
where $V(\phi)=U(R(\phi))/F(R(\phi))^2$. Note that the degrees of freedom in the field $g_{\mu\nu}$ in the original frame or 
Jordan frame (JF) split in the Einstein frame into a massless spin-2 field $\tilde{g}_{\mu\nu}$ and a massive scalar field $\phi$. 
Indeed, the action in this last frame is just the usual action in GR with an additional scalar field which propagates 
freely in the spacetime minimally coupled to gravity but nonminimally coupled to the matter fields. 
For simplicity, we work in the EF to solve the cosmological equations of motion and then we come back to the JF in order to 
interpret our results. We interpret the JF as the physical frame in which test particles move along geodesics and the energy momentum tensor of the matter fields is covariantly conserved. 

Recall that the metric in the JF is given by Eq.~(\ref{eq:metric}). 
The metric in the EF may be expressed as
\begin{equation}
d\tilde{s}^2=  F \left[ -dt^2 + a^2(t)d\textbf{x}^2 \right]  = 
-d\tilde{t}^2 + \tilde{a}^2(\tilde{t})d\textbf{x}^2\,  ,
\label{EFmetric}
\end{equation}     
where $d\tilde{t} = \sqrt{F}dt$ and $\tilde{a}=\sqrt{F}a$. The variation of the action, Eq.~(\ref{tensorscalaraction}), with respect to the scalar field $\phi$ and the metric 
 $\tilde{g}_{\mu\nu}$ result, respectively, in the following cosmological equations of motion~\cite{Faulkneretal:2006}:
\begin{align}
& 3\tilde{H}^2M_{pl}^2=\tilde{\rho}_M + \tilde{\rho}_{\phi} = \tilde{\rho}_M + \frac{1}{2}(\tilde{\partial}_{t}\phi)^2 + V(\phi)\, , \label{cosmologicalequationtwo}\\
& \tilde{\partial}_{t}^2\phi + 3{\tilde{H}}\tilde{\partial}_{t}\phi = 
-\frac{dV(\phi)}{d\phi} - \frac{\tilde{T}^M}{\sqrt{6}M_{pl}} = -\frac{\partial V_{\text{eff}}(\phi,\tilde{a})}{\partial \phi}\,,
\label{cosmologicalequationone}
\end{align}
with
\begin{equation}
V_{\text{eff}}(\phi, \tilde{a})= V(\phi) + \tilde{\rho}_M = \frac{3 \omega^2 M^2_{pl}}{4}\left( 1 - e^{-\frac{\sqrt{2/3}\phi}{M_{pl}}} \right)^2 + \tilde{\rho}_M.
\label{scalarpotential}
\end{equation}
Here $\tilde{H}$ is the Hubble parameter in the EF, and $\tilde{T}^M = -\tilde{\rho}_M + 3\tilde{p}_M$ is the trace of the 
energy momentum tensor of the matter fields in the EF, where $\tilde{\rho}_M$ and $\tilde{p}_M$ refer to the energy 
density and pressure, respectively. In addition, $V_{\text{eff}}(\phi, \tilde{a})$ is the effective potential acting on the scalar field, 
and $\omega = 1/\sqrt{3 a_2}$, as defined in Eq.~(\ref{relationomegaa2}).  
Let  $\tilde{\rho}_M$ consist of the usual energy density components of radiation ($\rho_{r}=\rho_{r,0}/a^4$ in the JF),
 nonrelativistic matter ($\rho_{m}=\rho_{m,0}/a^3$ in the JF), and
vacuum ($\rho_{\Lambda}$ in the JF), where $\rho_{r,0}$ and $\rho_{m,0}$ are the current energy density of radiation and nonrelativistic matter, respectively. Then, using the relation between both frames for the energy density $\tilde{\rho}_M=F^{-2}\rho_M(a)=F^{-2}\rho_M(F^{-1/2}\tilde{a})$, the effective potential can be written as
\begin{equation}
V_{\text{eff}}(\phi, \tilde{a})= V(\phi) + \bar{\rho}_r(\tilde{a})+\bar{\rho}_{m}(\tilde{a})e^{-\frac{\phi}{\sqrt{6}M_{pl}}} + \rho_{\Lambda} e^{\frac{-4\phi}{\sqrt{6}M_{pl}}}\,,
\label{effectivepotential}
\end{equation}
where $\bar{\rho}_m(\tilde{a}) = \rho_{m,0}/\tilde{a}^3$ and $\bar{\rho}_r(\tilde{a})= \rho_{r,0}/\tilde{a}^4$. 
\begin{figure}
\centering
\includegraphics[scale=0.7]{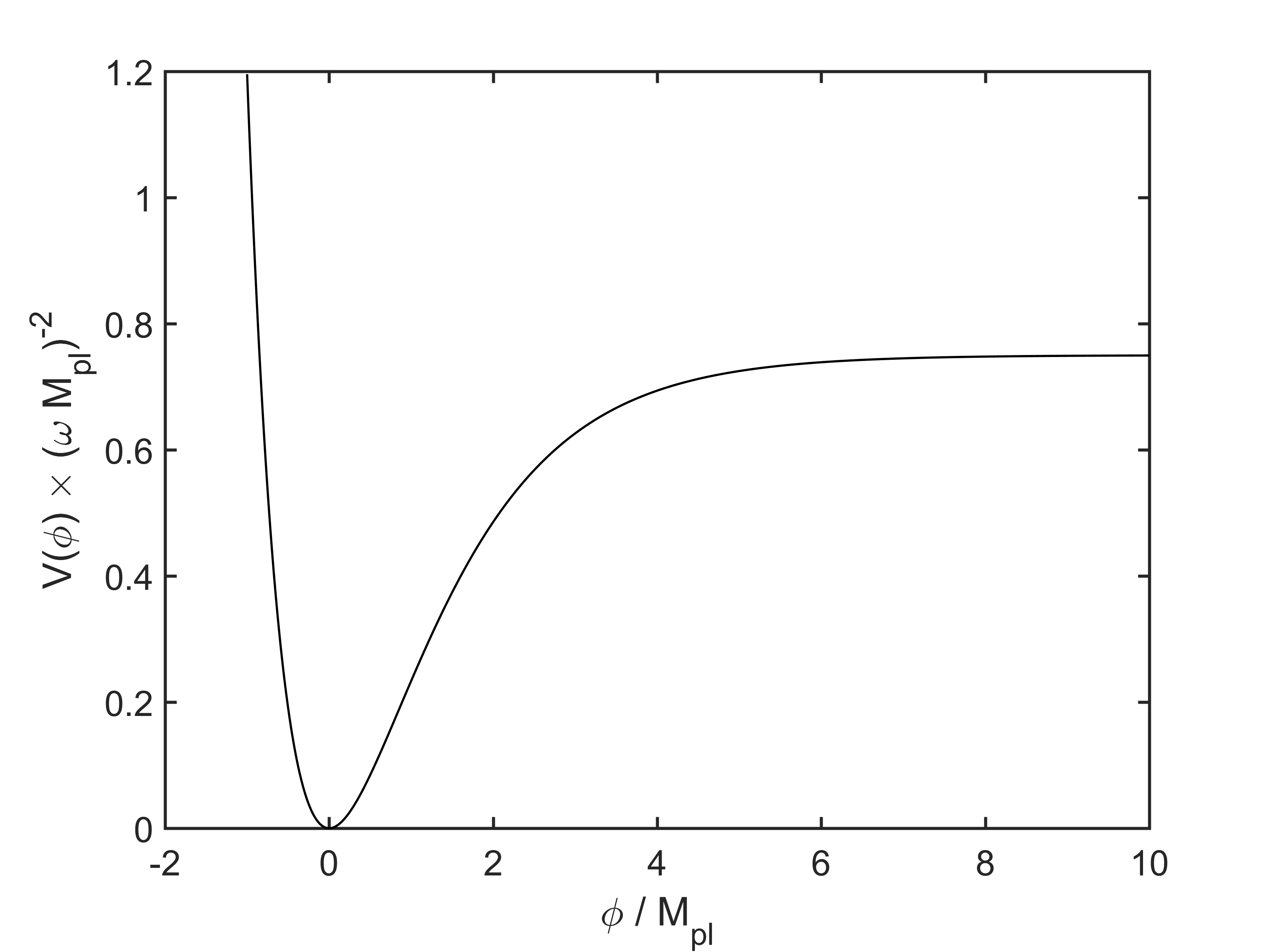}
\caption{The field potential $V(\phi)$, Eq.~(\ref{scalarpotential}), for the model $f(R)=R + (a_2/2) R^2$.}
\label{figure2}
\end{figure}

We analyze the cosmological effects of this model under the assumption of $|\phi|/M_{pl}\ll 1$ and in the regime 
$\tilde{H} \ll \omega$ where the oscillation time of the scalar field is much less than the expansion time in the EF.  
We do not treat $\phi$ as the inflaton field and neglect its possible decay in other particles.   

When $|\phi|/M_{pl} \ll 1$, the potential $V(\phi)$ shown in Fig.~\ref{figure2} can be approximated by a 
quadratic potential near the minimum at $\phi = 0$ leading to $V(\phi)\approx \omega^2 \phi^2/2$.  
Equation~(\ref{effectivepotential}) may also be expanded to write
\begin{equation}
V_{\text{eff}}(\phi, \tilde{a}) \approx \frac{1}{2}  \omega^2 \,\phi^2 + {\rm constant} + O(|\phi|/M_{pl} )\,.
\end{equation}
To leading order, Eq.~(\ref{cosmologicalequationone}) becomes
\begin{equation}
\tilde{\partial}_{t}^2\phi + 3\tilde{H}\tilde{\partial}_{t}\phi + \omega^2\phi = 0\,.
\label{scalarmotionequation}
\end{equation}
 The scalar field oscillates around the minimum of $V(\phi)$ with angular frequency $\omega$ and with 
 an amplitude that redshifts as $\tilde{a}^{-3/2}$ according to~\cite{Faulkneretal:2006} 
\begin{equation}
\phi({\tilde{t}})= \phi_i \left( \frac{\tilde{a}_i}{\tilde{a}} \right)^{3/2}\cos(\omega {\tilde{t}})\,,
\label{phioscillating}
\end{equation}
where $\phi_{i} > 0$ and $\tilde{a}_{i}$ corresponding to the oscillation amplitude and scale factor, respectively, when oscillations 
start at time $\tilde{t}_i$.    
The derivation of Eq.~(\ref{phioscillating}) can be obtained directly from the equation of motion of the scalar field. We will 
show this for the cases of a power law expansion and de Sitter spacetime in the EF, which are the cases of greatest
interest. Note that $\tilde{a}(\tilde{t})$ is a solution of Eq.~(\ref{cosmologicalequationtwo}). To leading order, the scalar
field energy density may be ignored compared to the matter contribution. This leads to the usual cosmological
solutions, such as  $\tilde{a}(\tilde{t}) \propto \tilde{t}^{1/2}$ for radiation, etc.

For a power law expansion, let $\tilde{a} \propto {\tilde{t}}^c$ with $c$ a constant. Then $\tilde{H} = c/{\tilde{t}}$ and 
Eq.~(\ref{scalarmotionequation}) becomes
\begin{equation}
\tilde{\partial}_{t}^2\phi + \frac{3c}{{\tilde{t}}}\tilde{\partial}_{t}\phi + \omega^2\phi = 0\,,
\label{powerlawA}
\end{equation}  
whose solution is
\begin{equation}
\phi({\tilde{t}})= {\tilde{t}}^{\frac{1-3c}{2}} \left[ C_1\,J_{\frac{3c-1}{2}}(\omega {\tilde{t}}) + 
C_2\, Y_{\frac{3c-1}{2}}(\omega {\tilde{t}})\right]\,,
\label{powerlawB}
\end{equation}
where $C_1$ and $C_2$ are constants and $J_\nu(z)$ and $Y_\nu(z)$ are the Bessel functions of the first and second kind, 
respectively.  The limit $\tilde{H}\ll\omega$ implies $\omega {\tilde{t}} \gg c$, 
and we assume that $c$ is of order one. Using the asymptotic forms of $J_\nu(z)$ and  $Y_\nu(z)$ for $\omega {\tilde{t}} \gg 1$, 
we have
\begin{align}
\phi({\tilde{t}}) & \propto {\tilde{t}}^{-3c/2}\cos(\omega {\tilde{t}})\,,\\
& \propto \tilde{a}^{-3/2}\cos(\omega {\tilde{t}})\,.
\label{powerlawC}
\end{align}
where we have ignored the phase in the argument of the cosine function.

For the case of de Sitter spacetime, let $\tilde{a} \propto \exp(\tilde{H}\, {\tilde{t}})$.
 Then  Eq.~(\ref{scalarmotionequation}) becomes
\begin{equation}
\tilde{\partial}_{t}^2\phi + 3 \tilde{H} \tilde{\partial}_{t}\phi + \omega^2\phi = 0\,.
\label{deSitterA}
\end{equation}  
Let $\phi({\tilde{t}}) \propto \exp(i\theta  {\tilde{t}})$ in Eq.~(\ref{deSitterA}), which leads to  
\begin{equation}
\theta^2 - 3i\tilde{H}\theta - \omega^2 = 0\,.
\label{deSitterB}
\end{equation}
The solution for $\theta$ is
\begin{equation}
\theta = \omega \sqrt{1-\left(\frac{3 \tilde{H}}{2\omega}\right)^2} + \frac{3}{2}i \tilde{H}\,,
\label{deSitterC}
\end{equation}
where we have selected the positive root. We can approximate Eq.~(\ref{deSitterC}) as
\begin{equation}
\theta \approx \omega  +  \frac{3}{2}i \tilde{H} + O\left( \frac{\tilde{H}^2}{\omega} \right)\,.
\label{deSitterD}
\end{equation}
Then the real solution of Eq.~\ref{deSitterA} has the form
\begin{align}
\phi(\tilde{t}) & \propto e^{-3 \tilde{H} \tilde{t}/2}\cos(\omega \tilde{t})\,,\\
& \propto \tilde{a}^{-3/2}\cos(\omega \tilde{t})\,.
\label{deSitterE}
\end{align}
Equations~(\ref{powerlawC}) and (\ref{deSitterE}) confirm the general expression, Eq.~(\ref{phioscillating}), for the
cases of primary interest. 

We may now combine Eq.~(\ref{auxiliarscalarfield}) with $\tilde{a} = \sqrt{F} a$ to write, under the condition 
$|\phi| \ll M_{pl}$,
\begin{equation}
a(t) = \frac{1}{\sqrt{F}}\, \tilde{a}(\tilde{t}) =   {\rm e}^{-\frac{\phi(\tilde{t})}{\sqrt{6}\, M_{pl}}} \; \tilde{a}(\tilde{t})
\approx  \tilde{a}(t) \, \left[1 - \frac{1}{\sqrt{6}\, M_{pl}}\, \phi(t) \right] \, ,
\end{equation}
where we have used $t \approx \tilde{t} + O(|\phi| /M_{pl})$.  Next we use Eq.~(\ref{phioscillating}) as
 $\phi(\tilde{t}) \approx \phi(t) = \phi_i (\tilde{a}_i/\tilde{a})^{3/2} \cos(\omega t)$ in this expression,  and then average 
 over the oscillations
to find that $\bar{a}(t) \approx \tilde{a}(t)$. The result may be written as Eq.~(\ref{scalefactorf(R)}). Note that the
assumption  $|\phi|/M_{pl} \leq \phi_i/M_{pl} \ll 1$ is equivalent to  $E_i \ll 1$ since $E_i \equiv [\phi_i/(\sqrt{6}M_{pl})].$ 

Now, note that in the regime $|\phi|/M_{pl} \ll 1$ and $\tilde{H}/\omega \ll 1$, the scalar energy density in the Einstein frame can be expressed as $\tilde{\rho}_{\phi} \approx (1/2)(\tilde{\partial}_t \phi)^2 + (\omega^2\phi^2/2) \approx (\omega^2 \phi_i^2/2) (\tilde{a}_i/\tilde{a})^3$ by using the expression for $\phi(\tilde{t})$ from Eq.~(\ref{phioscillating}). Then, the Friedmann equation in the Einstein frame, Eq.~(\ref{cosmologicalequationtwo}), can be expressed as
\begin{equation}
3\tilde{H}^2(\tilde{t})M_{pl}^2  = \left[ \frac{\rho_{r,0}}{\tilde{a}^4(\tilde{t})} \right] + \left[ \frac{\rho_{m,0}}{\tilde{a}^3(\tilde{t})} \right]e^{-\frac{\phi(\tilde{t})}{\sqrt{6}M_{pl}}} + \rho_{\Lambda}e^{-\frac{4\phi(\tilde{t})}{\sqrt{6}M_{pl}}}+ \frac{\omega^2 \phi_i^2}{2}\left[\frac{\tilde{a}_i}{\tilde{a}(\tilde{t})}  \right]^3\,.
\label{derivingFridemannequationstep1}
\end{equation}
Taking $t \approx \tilde{t} + O(|\phi| /M_{pl})$, $\tilde{a}(t) \approx \bar{a}(t)$, and Taylor expanding the exponential functions 
in Eq.~(\ref{derivingFridemannequationstep1}), we obtain
\begin{equation}
3\left[ \frac{1}{\bar{a}(t)}\frac{d\bar{a}(t)}{dt} \right]^2M_{pl}^2 \approx \left[ \frac{\rho_{r,0}}{\bar{a}^4(t)} \right]+ \left[ \frac{\rho_{m,0}}{\bar{a}^3(t)} \right] + \rho_{\Lambda} + \frac{\omega^2\phi_i^2}{2}\left[ \frac{\bar{a}_i}{\bar{a}(t)} \right]^3 + O(|\phi|/M_{pl})\,,
\label{derivingFridemannequationstep2}
\end{equation}
where $\bar{a}(t_i)=\bar{a}_i$. Note that Eq.~(\ref{derivingFridemannequationstep2}), in a rough  approximation, becomes 
 Eq.~(\ref{Friedmannequationapproximation}) when $\rho_{\phi} (t) < \rho_{m}(t)$.

\end{document}